\documentclass[compsoc]{IEEEtran}
\usepackage{graphicx} %
\usepackage{hyperref}
\usepackage[nameinlink]{cleveref}%
\usepackage[a4paper, margin=0.5in, footskip=0in]{geometry}%

\usepackage{makecell} %
\usepackage{booktabs} %
\usepackage{paralist} %
\usepackage[shortlabels]{enumitem} %
\usepackage{threeparttable}%
\usepackage{fancyvrb} \VerbatimFootnotes %

\newtheorem{protocol}{Protocol}

\usepackage{tocloft}
\usepackage{xcolor} %

\title{The Failure of Plagiarism Detection in Competitive Programming}
\author{
    \IEEEauthorblockN{Ethan Dickey}\\
    \IEEEauthorblockA{\textit{Department of Computer Science} \\
    \textit{Purdue University}\\
    West Lafayette, USA \\
    dickeye@purdue.edu}
}

\begin{document}

\maketitle
\begin{abstract}
    Plagiarism in programming courses remains a persistent challenge, especially in competitive programming contexts where assignments often have unique, known solutions. This paper examines why traditional code plagiarism detection methods frequently fail in these environments and explores the implications of emerging factors such as generative AI (genAI). Drawing on the author's experience teaching a Competitive Programming 1 (CP1) course over seven semesters at Purdue University (with $\approx 100$ students each term) and completely redesigning the CP1/2/3 course sequence, we provide an academically grounded analysis. We review literature on code plagiarism in computer science education, survey current detection tools (Moss, Kattis, etc.) and methods (manual review, code-authorship interviews), and analyze their strengths and limitations. Experience-based observations are presented to illustrate real-world detection failures and successes. We find that widely-used automated similarity checkers can be thwarted by simple code transformations or novel AI-generated code, while human-centric approaches like oral interviews, though effective, are labor-intensive. The paper concludes with opinions and preliminary recommendations for improving academic integrity in programming courses, advocating for a multi-faceted approach that combines improved detection algorithms, mastery-based learning techniques, and authentic assessment practices to better ensure code originality.
\end{abstract}

\setcounter{tocdepth}{2}
\tableofcontents
\section{Introduction}\label{sec:intro}
Academic integrity in programming courses is crucial both for fair assessment and for cultivating true coding competency. However, detecting plagiarism in source code has long been recognized as a difficult problem. Unlike textual plagiarism, plagiarized code can be easily masked by altering variable names, reordering logic, or inserting dummy code without changing functionality. In competitive programming (CP) courses - such as the CP1 course at Purdue University - the problem is exacerbated by the nature of assignments: many tasks have optimal or canonical solutions, increasing the likelihood that independent students produce similar code or that dishonest students copy existing solutions. Ensuring that each student's submission genuinely reflects their own work is challenging yet essential for mastery-based learning environments.

As a brief overview, CP is now widely regarded as a potent pedagogical instrument in computer science (CS) education. A growing body of empirical research shows that sustained engagement with CP sharpens students' algorithmic mastery, strengthens computational-thinking habits, and heightens overall problem-solving proficiency. Studies report that CP participation fosters greater learner initiative, lowers perceived difficulty of core programming concepts, and boosts retention in introductory courses \cite{bandeira2019teaching,picanco2018didactic}. Programming contests further cultivate self-directed learning, spur creative reasoning, and immerse participants in complex computational challenges \cite{raman2018students}. Longitudinal evidence from García and Aguirre (2014) demonstrates measurable skill gains attributable to continued contest involvement \cite{garcia2014learning}. More recently, Yuen et al. (2023) confirmed CP's practical educational value, showing that contest-driven experiences amplify student motivation, deepen independent learning and innovation, and better prepare graduates for advanced technical roles \cite{yuen2023computationalThinking}.

At Purdue, the introductory CP1 and CP2 classes use Kattis\footnote{\url{open.kattis.com}\label{footnote:kattis}} to manage weekly problem submissions.  The nature of such courses globally in the past and present is to assign high volumes of problems (at Purdue, four \textit{interview-style}\footnote{Historically, contest problems came first, then interviewers picked them up over time as good quick measures of programming ability.} problems a week) with (usually) no exams or quizzes, leading necessarily to the use of problem databases and common sources like Kattis or UVa Online Judge\footnote{\url{https://onlinejudge.org/}\label{footnote:uva}}.

Plagiarism in CS education is unfortunately common. Surveys of students have repeatedly shown a significant majority admit to some form of code copying or undue collaboration \cite{aasheim2012plagiarism}. For instance, in a large lower-division CS course of 200-300 students, teaching staff typically discovered 20-40 blatant cases of code plagiarism each semester \cite{chen2018cantstop}. These confirmed cases (around 10-15\% of the class) represent a lower bound, as instructors often focus only on the most obvious instances and ignore cases with plausible deniability. In other words, many subtle or well-disguised code copying incidents go undetected or ignored under current practices.  Other studies confirm this lower bound, with some as high as 75\% \cite{dickey2025evaluating,marsan2010why,fraser2014collaboration,joy2009taxonomy}. The true incidence of plagiarism is thus suspected to be higher, posing a serious threat to the fairness and educational validity of programming assessments and CS degrees.

To combat this, educators have turned to a variety of plagiarism detection tools and methods. Software tools such as Stanford's Moss (Measure of Software Similarity)\footnote{\url{https://theory.stanford.edu/~aiken/moss/}} have been in use since the mid-1990s and are considered a cornerstone for detecting copying among student programs \cite{schleimer2003winnowing}, and serve as the basis for other systems \cite{yan2018tmoss}. Online judge platforms like Kattis and UVa (\cref{footnote:kattis,footnote:uva}),%
widely used in competitive programming courses, now include built-in code similarity checks on every submission \cite{Olsson_Norlin_2020}. These automated systems can rapidly compare large numbers of submissions and flag suspiciously similar code, saving instructors considerable time. However, as we discuss later, they are far from foolproof - students can and do find ways to circumvent them. Moreover, similarity alone does not prove plagiarism: Moss's creators caution that it is \textit{``not a system for completely automatically detecting plagiarism''}, since it cannot judge intent or determine why two programs are similar \cite{schleimer2003winnowing}. Human judgment is still required to confirm actual cheating, and heavy reliance on raw similarity scores is a known misuse of such tools.

In response to the shortcomings of automated detectors, some courses (including our CP1/2/3 sequence) incorporate mastery-based learning techniques that emphasize genuine understanding. A notable practice is the use of regular interviews or viva voce exams\footnote{Traditionally thought of in the context of dissertation defenses, viva voce exams have a person defend their work in front of experts} with students about their code. In our experience, brief one-on-one code review interviews after assignment submissions provide a powerful authenticity check: if a student cannot explain the solution they turned in, describe where they struggled and how they solved their bugs, or reproduce key parts of it when prompted, it strongly indicates the work may not be entirely their own. This approach aligns with a broader pedagogical trend toward authentic assessment \cite{koh2017AuthenticAssessment}. Recent discourse suggests that oral exams can mitigate plagiarism and even counter the rising threat of AI-generated\footnote{Throughout this article, we use AI and genAI interchangeably. While generative AI is strictly a subset of AI, it is the main and by far most relevant type to academic dishonesty at the time of writing.} answers, since impromptu questioning is hard to fake \cite{dawson2025How}. Nonetheless, interviews are resource-intensive and impractical to scale to very large classes without significant qualified staff support.

Compounding these existing challenges, we now face a new paradigm: the advent of generative AI (genAI) tools like ChatGPT and GitHub Copilot\footnote{\url{https://openai.com/index/chatgpt/}, \url{https://github.com/features/copilot}}. These AI systems can produce correct or nearly-correct code from problem descriptions for a vast majority of first- and second-year computer science courses, enabling students to obtain working solutions without copying from peers or known sources \cite{lau2023BanIt,Becker2023programmingIsHard,Finnie22,Moradi23,dickey2024ai,hicke2023ai}. Early evidence shows this has already changed the landscape of academic dishonesty. A 2024 study observed that with ChatGPT's public release, the incidence of plagiarism in an introductory programming course significantly increased, as students shifted away from copying from online solution repositories (e.g. Chegg, CourseHero\footnote{\url{https://www.chegg.com/}, \url{https://www.coursehero.com/}}) toward using AI-generated code \cite{chen2024plagiarism}.  A 2025 study found that at least $75\%$ of computer science sophomores are using genAI for homework help, including students in a competitive programming course \cite{dickey2025evaluating}(Section 3.4).  Further studies on genAI in computing education can be found at \cite{raihan2025large,tanay2024exploratory,meyer2023chatgpt}.  Worryingly, conventional plagiarism detectors are ill-equipped to catch this form of cheating: if the AI-generated solution is original (not seen before in the class or a repository), it will not trigger similarity matches. Experiments confirm that code produced by ChatGPT often has lower pairwise similarity with other student submissions than students' codes do with each other, meaning it can slip through Moss undetected \cite{taylor2023plagiarism}. In short, genAI lets students ``cheat'' by outsourcing the coding work while largely evading current detection methods - a development that threatens to render traditional plagiarism policing methods obsolete if not addressed.

This was noticed \textit{before} the public release of ChatGPT by Biderman and Raff \cite{biderman2022fooling}, among others \cite{drori2022neuralnetmath,drori2021solvinglinearalgebraprogram,shporer2023learning,tran2021solving}, who showed that students using GPT-J \cite{wang2021gptJ} could evade MOSS detection.  For traceability, the origins of pre-GPT-4-style pretrained transformers (which genAI is) are nicely laid out in \cite{gao2020pile}.  

In this paper, we offer a preliminary analysis of the shortcomings of current plagiarism detection methods in competitive programming courses and outline possible avenues for improvement.
Drawing on both the academic literature and the author’s seven semesters of experience as a CP course instructor and ICPC team coach - underpinned by a formal background in higher-education pedagogy and education research - our goal is to bridge theory and practice in pursuit of more effective detection strategies.
The next section provides background and related work, including definitions of code plagiarism and an overview of detection techniques from the literature. We then share experience-based observations from the CP1 course, highlighting concrete examples of plagiarism attempts and detection outcomes. Following that, we offer an analysis of current detection methods - Moss, Kattis, manual code review, and suspect interviews - examining how each works and where it falls short in the competitive programming context. A comparative summary table of these methods' strengths and weaknesses is included in \Cref{tab:summary_strengths_weaknesses}. Finally, in the discussion, we offer opinions and suggestions for improvement, advocating for a multi-faceted strategy (combining enhanced tools, process changes, and educational practices) to better uphold academic integrity in programming courses. Through this rigorous reflection, we aim to inform computer science educators and researchers seeking robust solutions to the evolving challenge of code plagiarism.

\section{Background and Literature Review}\label{sec:backgroundandlit}
\subsection{Defining Code Plagiarism}
Source code plagiarism is typically defined as the act of presenting someone else's code (in whole or part) as one's own work, without proper attribution, in a context where originality is expected. Parker and Hamblen's classic definition (1989) describes software plagiarism as \textit{``a program which has been produced from another program with a small number of routine transformations''} \cite{parker1989computer}. In other words, plagiarism may involve copying code and then applying minor edits to conceal the copying. Researchers have long studied the spectrum of obfuscations a cheating student might perform. Faidhi and Robinson (1987) proposed a six-level taxonomy of program modifications to characterize degrees of plagiarism \cite{faidhi1987empirical}. At the lowest level (Level 1), a student only changes comments in the code; higher levels involve renaming identifiers (Level 2), altering declarations (Level 3), inserting redundant statements or reordering procedures (Level 4), changing control structures like conditionals (Level 5), and finally, at Level 6, making substantive changes to the program's logic or algorithm \cite{mirza2015style,mirza2018style}. Importantly, each higher level subsumes the previous - e.g. a Level 4 plagiarism will typically also include changes to identifiers and comments. By Level 6, the code might be logically distinct despite solving the same problem, making it extremely difficult to detect through surface similarity alone (and hard to distinguish from sources like TA assistance).  Other perspectives on types of plagiarism can be found in \cite{joy2009taxonomy, dick2002addressing, fraser2014collaboration}; an excellent systematic literature review on plagiarism can be found in \cite{albluwi2019plagiarism}.

We introduce a new term useful for comparison to direct plagiarism: \textit{Whiteboard similarity} is similarity in implementation ideas between two students due to completing the problem-solving process together, but without seeing each other's implementations.  This kind of process is usually allowed in computer science courses, especially as difficulty of course material increases.  As opposed to code plagiarism, whiteboard similarity will almost never result in plagiarized code, or even code that gets flagged as plagiarism, except when the answers are short (as discussed in the limitations of automated detection software later in this section).  From 4 years of reviewing 100 students' code over 40 problems across a semester, even two students who consistently work together will have drastically different code fingerprints (see discussion on code stylometry in \Cref{sec:background:genai}), so long as they did not look at each other's code.  This is in direct contrast with an idea like \textit{code collaboration}, which directly results in plagiarized code (presuming independent submissions are required).

\subsection{Prevalence and Academic Impact}
Plagiarism and excessive collaboration in programming courses are widely acknowledged problems. A survey by Mirza and Joy (2017) across 18 UK universities found a majority of students were aware of code plagiarism issues and some had participated in unauthorized code sharing \cite{mirza2017style,mirza2015style}. Studies consistently indicate that a non-trivial percentage of students engage in these practices, undermining learning outcomes. From an academic integrity perspective, plagiarism devalues honest students' efforts and can skew grading fairness \cite{chen2018cantstop}. From a learning perspective, students who plagiarize code bypass the essential practice and struggle that leads to mastery, often resulting in weaker understanding. Recent research has empirically demonstrated this learning deficit: in a large introductory course, Chen et al. (2024) showed that students who frequently submitted plagiarized or AI-generated solutions saw significantly lower exam performance later, with learning losses proportional to the amount of copying \cite{chen2024plagiarism}. In essence, cheating may ``work'' to get assignment points, but it leaves the student ill-prepared for assessments that require true understanding, echoing the adage that one only cheats oneself in the end.

\subsection{Automated Code Similarity Detectors}
To identify potential plagiarism, educators have deployed various automated tools that measure similarity between programs. The most famous is Moss, originally developed by Alex Aiken in 1994 \cite{sanders1997online}. Moss automatically compares all pairs of submissions in a class and computes a similarity score for each pair, highlighting matching code segments \cite{schleimer2003winnowing}. It uses a robust winnowing algorithm for document fingerprinting to detect similar subsequences even if students have renamed variables or slightly reordered code. Moss and similar systems (such as JPlag, developed in 1996 \cite{sauglam2024obfuscation}) have been very effective at catching straightforward cases of copying and even moderately transformed code. These tools are credited with greatly reducing the effort needed to uncover plagiarism - instead of manually performing $N^2$ comparisons, instructors get an itemized report of the most suspicious pairs (see \Cref{app:MOSS} for examples). Consequently, Moss has become a standard in CS education worldwide for plagiarism detection \cite{luke2014software,sheahen2016taps}.

However, automated similarity detectors have well-documented limitations. One issue is false positives on short or simple assignments. When an assignment has only a few plausible solution approaches or very basic code (common in introductory or competitive programming exercises), many students' independent solutions will naturally look alike. In such cases, Moss can report high similarity across many submissions even when no collusion occurred. Hicks et al. (2024) describe this phenomenon in an analysis of a programming exercise dataset: for some short problems ($\approx 10$ lines of code solutions), Moss flagged 93\% of submissions as similar, %
despite no evidence of cheating, simply because the optimal solutions converged on the same structure. Setting an appropriate similarity threshold becomes unreasonable in these situations, as even honest work can appear suspicious \cite{hicks2024approach}. Instructors must then manually distinguish legitimate similarity from actual copying - a non-trivial task.

Another limitation is that classical detectors can be fooled by systematic code transformations beyond their detection capability. While Moss and JPlag are resilient to superficial changes (e.g. Level 2 identifier renaming), more complex obfuscations can evade detection. Students intent on cheating have access to ``plagiarism laundering'' tools that automatically transform code. A striking example is a tool called Mossad (a darkly humorous nod to Moss) which repeatedly inserts dummy statements into a plagiarized program without changing its behavior \cite{devore2020mossad}. By bloating and rearranging code, such a tool can reduce the measurable similarity between the plagiarized code and the original source, potentially slipping under Moss's radar. Indeed, the emergence of automated obfuscation attacks is challenging the old assumption that ``it takes more effort to evade the detector than to write the code from scratch'' \cite{sauglam2024obfuscation}. Research is ongoing to improve detectors' robustness. For example, a recent update to JPlag (v5.0) incorporates token-sequence normalization and syntax-tree analysis to better catch reordered or inserted code sequences. These enhancements reportedly allow JPlag to significantly outperform earlier tools (including Moss) in detecting heavily obfuscated plagiarism \cite{sauglam2024obfuscation}. Nonetheless, no detector can guarantee catching all cheating - there is a continual arms race between plagiarism tactics and detection techniques. 

\subsection{Integrated Online Judge Systems}
In competitive programming education, it's common to use online judges to automatically grade solutions for correctness and performance. Some judge systems also integrate plagiarism checks. Kattis, an automated coding assessment platform popular in ICPC training and CP courses, is one such system. At KTH (Royal Institute of Technology), Kattis is used for student assignments and it \textit{``runs plagiarism detection on every submission against every previous submission to the service''}, with results available to teaching staff \cite{Olsson_Norlin_2020}. Kattis's plagiarism detector employs a combination of hash-based comparisons and n-gram analysis of code to find similarities. The advantage of this integration is twofold: plagiarism checks are continuous (every new submission is instantly checked against a large database of past solutions, possibly including those from prior semesters, other institutions on the platform, and published solutions as on GitHub), and the process is automated within the grading workflow. Instructors are alerted to suspect cases without needing to run a separate tool like Moss manually. However, Kattis's approach shares similar underlying principles and limitations as standalone tools. Hash and n-gram comparisons can catch identical or slightly modified code, but significant refactoring or logically equivalent reimplementation can elude detection. Additionally, because Kattis casts a wide net (comparing to all submissions in its database), it may sometimes flag students who independently wrote similar code for a well-known problem - again requiring human adjudication to distinguish coincidence from copying.

A well-known but rarely discussed issue is the publishing of solutions to online judge systems.  Kattis itself requests users not to publish code online\footnote{\href{https://support.kattis.com/support/solutions/articles/79000112016-can-i-publish-my-solutions-online}{Kattis FAQ: Can I publish my solutions online?}}, but acknowledges that users own the code they write.  This causes issues in problem difficulty calculations and more relevantly, assessment of honest problem completion.  We include a non-comprehensive list of Kattis solution GitHub repositories in \Cref{app:trackedRepos}.  We include solutions from all of these repositories plus several variations on LLM-based solutions when running plagiarism detection.

\subsection{Manual Code Review and Visual Inspection}
Long before automated tools, instructors relied on manually reading through student programs to identify cheating. Experienced educators can often spot telltale signs of plagiarism by eye. For example, if two students who supposedly worked independently produce source files with the same unusual indenting or the same atypical variable names and comments, an instructor's suspicion is naturally raised. Unique logical quirks or bugs replicated across submissions are another giveaway - if two assignments have the same subtle flaw in output formatting or the same creative but non-obvious algorithmic approach, one might be a copy of the other. Manual review can also incorporate knowledge of the students: an instructor might notice if a weaker student suddenly submits an exceptionally polished solution out of character for their ability. In such cases, even if an automated tool did not flag the code, the instructor's intuition might.

The strength of manual inspection lies in its flexibility and context-awareness. A human can consider semantics and intent, not just text similarity. Manual review can catch cases that confound automated detectors (for instance, code that is functionally the same but written in a completely different style might evade Moss but an expert could recognize the identical solution strategy). It also helps filter out false positives: as Moss's documentation stresses, a human must ultimately decide if a similarity is innocuous or indicative of cheating \cite{schleimer2003winnowing}. In practice, course staff often use automated tools to narrow down candidates, then manually compare those code pairs side-by-side to make a judgment \cite{chen2018cantstop}.

The obvious weakness of manual review is scalability. Reading through potentially hundreds of programs (or diffing dozens of pairs) in a large class is extremely time-consuming. Chen (2018) reports an instructive anecdote: in a class of $\approx 250$ students, an initial Moss run might flag around 1000 pairwise comparisons to examine, which the TAs then manually \textit{winnow} down to $\approx 100$ suspicious cases and finally 20-40 confirmed cheats \cite{chen2018cantstop}\footnote{Moss uses a winnowing algorithm.}. This represents a massive effort in human hours. Moreover, manual detection is inherently subjective and prone to inconsistencies; it depends on the vigilance and experience of the reviewer. Two different instructors might not always catch or treat a borderline case the same way. Thus, while indispensable, manual review is best used in conjunction with tools and for confirming, rather than as the primary detection mechanism in large courses.

\subsection{Interviews and Oral Assessments}
An alternative approach to detecting plagiarism - one that shifts from analyzing the submitted code to probing the student's understanding - is the code interview. In academia, this is akin to a viva voce exam for programming assignments. After a student submits their code, an instructor or TA meets with them (in person or virtually) and asks them to explain how their solution works, why they chose certain implementations, or even to write or modify a portion of the code on the spot. The premise is simple: if the code is truly the student's own work, they should be able to comfortably discuss its design and details; if it was copied or generated without genuine understanding, this should become apparent in their inability to explain it clearly. Even a very well-disguised plagiarism can be uncovered if the student cannot answer specific questions about ``their'' code.

Our CP1 course uses regular interviews as part of a mastery-based learning model - every student periodically has to walk through their recent solutions with an instructor. This has proven to be a powerful deterrent against plagiarism. Students know that copying a solution without learning it is futile, because any short-term gain on the assignment will be negated by the subsequent interview where they must demonstrate mastery. From a detection standpoint, interviews have high effectiveness: they directly address authorship in a way no static analysis can. Indeed, even some students suggest this method - when faced with an accusation, an honest user on an online forum insisted ``tell them to ask you about anything in your code, and answer it. It's easy to copy, but hard for people to fully understand code they copied'' \cite{anonymousprofessor99Reddit}. This insight encapsulates why an oral exam can catch plagiarism that tools miss.

The downsides, however, are significant. Conducting individual interviews for an entire class is labor-intensive. For a class of 100, even a 10-minute interview per student is 1000 minutes (over 16 hours) of instructor time for one round of interviews. Our implementation manages this by being selective (e.g. focusing on key assignments or higher-risk students) and by integrating the interviews into the course structure so that they replace some other grading tasks. Still, scalability is a concern - not every institution has the resources to do this routinely. Additionally, the success of an interview depends on the skill of the interviewer to ask probing questions and the integrity of the process (a student might prepare rehearsed explanations for copied code, though follow-up questions can usually pierce that facade). There is also a psychological aspect: some students experience anxiety in oral exams, and care must be taken to conduct interviews in a fair, consistent manner for all. Despite these challenges, the literature on academic integrity increasingly notes viva voce assessment as a promising practice, especially in the age of AI. As one commentary put it, oral assessments are an ``interesting alternative'' because a student's spoken responses and on-the-fly problem-solving can't be easily produced by a chatbot or someone else \cite{dawson2025How}. In summary, interviews are perhaps the most direct way to verify code authorship and understanding, but they require considerable effort and pedagogical commitment to implement at scale.  We provide our interview protocol in \Cref{protocol:interviews}.

\subsection{GenAI and the New Plagiarism Landscape}\label{sec:background:genai}
The rise of AI code generators (ChatGPT, Codex/Copilot, etc.) in the past three years has introduced a fundamentally new challenge for plagiarism detection. Previously, most code plagiarism involved a human source - a classmate, an online repository, or solution banks - meaning there was a tangible original that could potentially be found or matched. Now, a student can prompt ChatGPT to solve an assignment; the AI may produce a correct (or nearly correct) solution that is novel (not directly copied from any existing code, but generated by its internal model). To plagiarism detectors, this AI-generated code appears as the student's original work, unique across the class. Early research confirms that AI-generated code does not set off alarms in the way traditional copying does. In a controlled experiment, ChatGPT-produced solutions had measurably lower similarity to other students' code than students' solutions did to each other, when analyzed by Moss \cite{taylor2023plagiarism}. In a mixed pool of files, the ChatGPT files blended in and were not obvious outliers by Moss's metrics. The researchers concluded that \textit{``even if ChatGPT has been used in the past few months for coding assignments, it may not have been detected reliably by plagiarism-checking software''}. This aligns with a 2022 study that showed that AlphaCode, a code generation system by DeepMind based on the same principles that LLMs are \cite{li2022alphaCode}, has an average maximum similarity score of 0.56 (out of 1) with human-written code (on difficult problems, higher for simpler ones) and that the tool performs about the same as the human code in terms of runtime and memory usage \cite{lertbanjongngam2022empirical}.
Both of these results align with anecdotal reports from instructors: plagiarism tools are catching fewer cheating cases, yet suspect code may be on the rise - suggesting students are exploiting AI help which leaves no easily detectable trace.

That is not to say AI-generated code is impossible to detect, but it requires new tactics. Some AI outputs have quirky telltale signs (for example, earlier versions of ChatGPT sometimes included unnecessary helper functions or odd naming conventions). A ChatGPT plagiarism study noted that some generations contained ``strange issues, like referencing undefined methods, [which] would be a red flag for AI-generated code'' if submitted as-is \cite{taylor2023plagiarism}. However, a savvy student will likely test and fix such issues before turning in the code. Another idea is to look at similarity trends rather than absolute values. If an instructor has a large set of submissions, and most student-student comparisons cluster around a certain similarity range, an AI-generated submission might show a pattern of being uniformly low-similarity to everyone. In theory, this could raise suspicions (``one of these things is not like the others''). The aforementioned experiment did observe that student vs. student code comparisons often showed more overlapping lines than student vs. ChatGPT comparisons \cite{taylor2023plagiarism}. In ideal conditions, a teacher with an ``eye for details'' might notice that one submission strangely has very little in common with any other, sticking out not due to similarity but dissimilarity. This is a very subtle indicator, though, and not a practical general solution.

At present, mainstream plagiarism detectors (Moss, etc.) cannot reliably identify AI-generated code as plagiarism, since by definition plagiarism detection looks for matches with known sources. AI output is essentially a new, unseen source. Detecting AI assistance likely requires different approaches - possibly stylometric analysis (i.e., identifying the author of code by style, and detecting when a submission's style diverges from the student's known coding style) \cite{caliskan2015anonymizing}, or specialized AI detectors (analogous to GPT essay detectors, though those have proven unreliable so far \cite{gritsai2025aidetectorsgoodenough,lin2024detectingmultimediageneratedlarge}). The academic community is actively researching this; for example, there are studies on using machine learning to attribute code to authors based on patterns in how code is written \cite{mirza2015style,mirza2018style}. In concept, such techniques could flag a submission that doesn't match the profile of the student's earlier work (hinting it was produced by someone or something else). Until these methods mature, instructors are adapting in other ways, like designing assessments that are harder for AI or instituting oral defenses. The genAI revolution is a double-edged sword: it offers new tools for learning but also new avenues for misconduct. Our focus here is on the latter - how it undercuts traditional detection - and it will factor into our analysis and recommendations.

In summary, the literature reveals a multifaceted problem: (1) plagiarism in programming can take many forms, from trivial copy-paste to highly convoluted code theft, (2) our detection arsenal (Moss, Kattis, JPlag, etc.) is effective against some forms but can be outmaneuvered by determined cheaters or novel AI helpers and (3) purely manual or oral methods, while effective, struggle to scale. With this background, we now turn to the specific context of competitive programming courses and the author's experience to ground the discussion in real instructional practice.

\section{Experience-Based Observations from a Competitive Programming Course}
The insights of this paper are informed by the author's experience as Instructor of Record for Purdue University's Competitive Programming I (CS211: CP1) course over seven semesters (spanning 2022-2025), and as the lead designer for the official Purdue CP1/CP2/CP3 course sequence. Each semester, CP1 enrolls roughly 100 students (120 if one counts drops in the first 2 weeks), primarily second- and third-year Computer Science majors, and introduces them to algorithmic problem solving in an online-judge setting. The course is run for 12 weeks\footnote{A 12-week course with 2 credit hours is approximately equivalent to a 3-credit hour workload for 2/3 of the semester.} in a mastery-based learning style: students are expected to fully solve a certain number of programming problems each week, with the opportunity to fix and resubmit solutions before the weekly deadline until they pass all test cases. Rather than traditional exams, assessment is continuous and based on achieving problem-solving milestones. This structure inherently reduces the pressure to cheat on any single assignment - since students can try again after failure and are encouraged to learn from mistakes, there is less incentive to present someone else's work as your own just to get a one-time grade. In theory, mastery learning should promote honesty by focusing on learning progress over high-stakes performance.  In CP1, we have a non-standard application of mastery-based learning, and thus promotion of honesty is more nuanced.

Nonetheless, plagiarism has not been eliminated. Over multiple offerings of CP1, a pattern of cheating behavior became evident. A minority of students each semester (on the order of 10-20 students, $\approx 10-20\%$ of the class) still attempted to shortcut the process by submitting code that was not their own original effort. Some tried the classic route of copying a solution from a classmate or friend. Given that CP1 problems are administered on Kattis, many others resorted to searching online for existing solutions - usually from GitHub (see \Cref{app:trackedRepos}) - and either copying it directly (easy to detect with Kattis, usually when the student became frustrated their adaptation of the solution didn't work) or by morphing their code to match the solution (much harder to detect). Since S22, with the rise of GitHub Copilot first and later ChatGPT, we observed a new phenomenon: an increase in submissions that had unusual coding style or constructs that did not match the student's previous work. In \textit{\textbf{many}} instances, students directly admitted (when confronted) that they had used ChatGPT or a similar tool to generate a solution, thinking it was somehow ``safer'' (less likely to be caught) than copying from GitHub or peers.  Since F22 (before ChatGPT was publicly released), we have had a statement like the following in the syllabus:
\begin{verbatim}
Warning regarding GitHub Copilot and other
automatic code-generation software:

If your submission is flagged as similar or
identical to other submissions on Kattis and
it is because you used GitHub Copilot or
other automatic code generation software,
this will count as a violation of the
academic integrity policy. By using
automatic code generation, you are switching
from a creative process to a review-based
process. It becomes much harder to spot
errors _and_ prevents you from learning the
material to the standards required.
\end{verbatim}

\subsubsection*{Aside: Plagiarism Policies}
This is also covered (with much less commentary) in our parallel paper \cite{dickey2025removing}. Policies regarding assessment and plagiarism have evolved every semester since the courses were rewritten in Spring 2022 (S22).  Generally, CP1 and CP2 are structured to require students to complete an average of 3.5 problems a week including bonus credit, with one point awarded per problem completed.  Four problems as week are available, and one problem a week is denoted the ``\textit{pair problem}'' - an easier problem designed to be done in the last 20 minutes of one lecture a week with an assigned partner.  In terms of assessment, weekly interviews were implemented starting in S25.  10 students a week ($\approx 10\%$) were called into Instructor or Head TA office hours to have an ``interview'' over a randomly selected problem they completed the week prior (announced on Monday the week they were required to come in).  The only constraint on the randomness was that every student was selected at least once.  With 12 topics, at most 20 students were selected twice and a few were selected 3 times.  The intention was to provide some sort of authentic assessment of student ability to provide both some extrinsic motivation for complete understanding of submitted code and some guarantee to the instructor that students were learning the course material, since there are no other in-person assessments (as is standard in CP course design).

In S25, the plagiarism procedure was as follows:
\begin{enumerate}[(1)]\label{list:plag_procedure}
    \item Detection of plagiarism by MOSS (similarity scores rising above the average and explicitly ignoring similarity due to common code bases like FastReader\footnote{Something like this is provided to students at the start of the semester: \href{https://www.geeksforgeeks.org/fast-io-in-java-in-competitive-programming/}{Geeks for Geeks Fast IO}}), Kattis auto-flagging, TA identification (usually when students brought suspicious code to office hours for help), interviews, or manual review (things such as one-shot correct answers, no submissions until the last 1-2 hours, rapid code changes over a few minutes, language changes - mostly manual review of code sylometry).
    \begin{itemize}
        \item One TA was assigned the responsibility of weekly MOSS runs and was not expected to complete any other course-related tasks.  They often had bandwidth to pick up other tasks, but MOSS runs consumed on average 1hr per problem, or 4 hours per week for just the initial reports.  Each problem averaged 80 last accepted (student-unique) submissions.  Verifying, emailing, and recording MOSS incidents took the instructor 5-10 minutes per case, resulting in another 40 minutes of work per week, on average, in Spring 2025, not including the extreme cases (see \Cref{tab:num_plagiarism_incidents} and description later in this section).
        \item Our policy on what threshold of plagiarism we are willing to pursue is defined as follows: we do not attempt to detect a student who puts in more effort to cheat than actually doing the problem.  Almost certainly, these students have, to some degree, met the basic Learning Objectives of the course because they must understand the code they are copying in order to modify it to such a high degree.  The thought behind this is that this process comes too close to students who have whiteboard similarity -- those who worked together -- to be reasonably separated from the noise. We attempt to detect all those who put in less effort and understanding than this category of students to plagiarize.
    \end{itemize}
    See \Cref{app:MOSS} for further analysis and examples of MOSS runs in CP1.
    
    \item Manual review by instructor for every case to confirm suspicion; in 80\% of cases, the code MOSS flagged as similar was easy to spot as identically (re)written; in the next 19\% of cases, manual review was able to clearly identify rewritten solutions; in the final 1\% of cases, it was a genuine false positive.  This distribution can be explained by students not being very creative with their copying.
    \item After plagiarism is confirmed, the instructor sends the student an email that says something like ``You have been flagged for plagiarism on X problems on topic Y.  You will receive -2 per plagiarized problem.  You are also required to take the final.  The next time we catch you it will be an automatic report and fail.'' with the subject line ``[No Response Required] CP1...''  Each problem is worth 1 point, so this results in a net score of -1/problem on their grade.
    \item If students protest, the instructor usually responds with something similar to ``Your code for Z matched [ChatGPT, a GitHub Solution, etc.] identically.'' and the student usually does not respond, or apologizes if they did.  Interviews/meetings are rarely required.

    \item If catching a student a second time, a meeting always occurs.  From that meeting, usually a formal report is submitted and a plan is made for the student to submit makeup problems for each plagiarized problem in order to pass the course.
\end{enumerate}
The statement about taking a required final is new as of S25.  Traditionally, a 5-hour solo bonus contest is held at the end of the last topic, with 7 random problems that cover CP1 material.  Up to 7 bonus points can be gained from this contest, but like any well-designed ICPC-style contest, no one should be able to finish all of the problems while every problem gets solved and every team (of 1 here) solves at least 1.  This is the first and only experience CP1 and 2 students get with a timed contest setting\footnote{See parallel course design paper on removing the competition from CP1 and 2 \cite{dickey2025removing} and sister paper on putting competition back in CP3 \cite{luo2025curriculum}.}.  The ``final'' was completion of 3 problems during the (in-person) bonus contest.  A number of students who were required to take the final needed extra time (up to 1-1.5 hours longer than the 5 hour limit) and limited TA assistance to finish.  This was allowed as this (S25) was the first iteration of such a policy.  We provide a brief overview of the number of plagiarism incidents recorded in \Cref{tab:num_plagiarism_incidents}.  In S24, we began using MOSS, contributing to a drastic rise in plagiarism cases.  In F24, for reasons we do not report here (as part of an experiment, see \cite{dickey2025evaluating}), students were allowed to use genAI however they wanted for topics 7-11 (the last $\approx 41\%$ of the course), with significantly harder problems assigned (though not hard enough to escape genAI while being CP1-level).  In S25, we restricted this intervention-based-privilege to a single week, topic 7.

\begin{table}
    \centering
    \begin{tabular}{c|c|c|c|c|c|c|c}%
        Semester & S22 & F22 & S23 & F23 & S24* & F24 & S25\\
        \midrule
        \shortstack{\# Incidents,\\\# Serious} & \shortstack{8, 0\\~} & \shortstack{6, 1\\~} & \shortstack{13, 2\\~} & \shortstack{19, 4\\~} & \shortstack{31, 0\\~} & \shortstack{18, 0\\~} & \shortstack{46, 16\\~}\\
        \bottomrule
    \end{tabular} \vspace{1mm}%
    \caption{Number of (confirmed) plagiarism incidents per semester. Serious incidents are defined as students caught plagiarizing on 4 or more problems or students caught plagiarizing a second time after being notified.\\~*MOSS runs began in S24.}
    \label{tab:num_plagiarism_incidents}
\end{table}

\subsubsection*{Back to General Observations: Major and Minor Examples}\label{sec:experiences:backtogeneralobsv}
One real example highlights the failure of automated detection in our context.  In S25, no less than 15 students submitted a Binary Search The Answer (BSTA) solution to Topic 2: Greedy.  This was immediately a red flag for TAs in office hours as (1) we don't teach BSTA until topic 4 and (2) no other CS class at this university teaches BSTA.  Many teach Binary Search, but BSTA is a highly niche and unintuitive problem-solving technique where a program binary searches through a (BSTA = boolean, Bisection = continuous) monotonic \textit{answer space} (such as whether two opposing monks have passed each other on a mountain\footnote{\url{https://open.kattis.com/problems/monk}}) using an efficient \textit{can} function (\textit{can(time)} you tell me if the monks have passed each other at \textit{time} efficiently?).  After having two separate TA reports, we manually checked through every accepted submission and found 15 total students who had submitted using a more advanced technique that they had no reason to know
\footnote{We have also caught many students in the past using a min/max monotonic queue on \href{https://open.kattis.com/problems/sound}{The Sound of Silence}, where the intended solution used a simple balanced Binary Search Tree (i.e. red-black tree: c++ map, Java Treemap, Python doesn't have one) that keeps track of the min and max with $\log$ time random element removal. This use of a bBST in replacement of a priority queue for a similar purpose was explicitly covered in the classes before the deadline. Since Python doesn't have a bBST, students using python cheat to figure it out, find the monotonic queue solution, fail to understand it, and copy it into their code.  These are trivial to spot as the code is frequently much more elegant and simple than we expect to see from students.}.
Dealing with these cases was a nightmare\footnote{In the end, 11 students were caught and due to the blatant disregard for even understanding the code they submitted, all students were reported in one mass report to the university, despite normal policy allowing one strike before reporting.}. 
To determine if a student truly learned what a monotonic queue was and implemented it on their own (which was true of 4 students of the 15, all notably high-performers), which was possible due to ``whiteboard collaboration'' being allowed, we interviewed each one.  We used a modified version of the interview protocol below with an additional statement at the beginning similar to that of the plagiarism procedure in the above \textit{Aside: Plagiarism Policies}.  The Head TA for CP1 attended these meetings and ran some of them.

~\hrule
\begin{protocol}[Interview Protocol (Standard)]\label{protocol:interviews}
    Assumes access to the problem statement but \textit{no} access to their code.  Instructor is expected to know the problem well and have their solution pulled up to verify the student explanation matches their code.
    \begin{enumerate}
        \item \textit{What is the problem asking you to do?}
        \item \textit{Explain [a sample input].}
        \item \textit{Explain your approach to this problem.}
        \begin{enumerate}[a]
            \item During their explanation, they need to hit all key insights within a CP1 understanding.
            \item They need to explain the output if it's complex.
            \item They need to be able to explain what paradigm or approach with CP1 terminology (connect back to CP1 terminology from class, etc.). This is a looser requirement as long as they can describe the approach using similar language.
        \end{enumerate}
        \item \textit{Why is your solution fast enough?}  They should be able to explain that we get about $1e9$ operations (accounting for overhead) and their solution's runtime with the given input size is $\le 1e9$.
        \item~[Optional, use if plagiarism suspected but not confirmed yet.]  \textit{Where is somewhere you struggled/what is a bug you had and how did you fix it?}  Students who took a shortcut in learning are \textit{very} rarely able to answer this (and the ones who do are usually fast enough learners that they understand the solution well enough to meet most learning objectives anyways), so we label this our silver bullet.  Students will likely need to warm up with the other questions first, however.
    \end{enumerate}
    If all else fails and you have evidence of plagiarism (e.g. other students' confessions for highly similar code), you can ask, \textit{Please open ChatGPT on your laptop in front of me.} and look through their history.  This is a sledgehammer and destroys any semblance of teacher-student trust, so should only be used as a last resort.

    \noindent If an undergrad is giving the interview, and they have reached the end and determined the student does not know their code well enough to pass, they say something like \textit{Thank you for coming in, we will get back to you later this week on your grade.}  Undergrads should never be required to fail other undergrads based solely on their own evaluation and interaction.  They are usually correct, but this creates odd power dynamics and misplaced responsibilities.  The instructor usually does not meet with failed students.
\vspace{1mm}\hrule
\end{protocol}

We only had to use ``open ChatGPT in front of me'' two to three times out of the hundreds of plagiarism interviews we've conducted.  In fact, only during the 11 plagiarism cases on Topic 2 of S25.  One would expect MOSS or similar tool to catch these cases easily.  However, due to the students' use of ChatGPT and subsequent manual implementation, few of them were flagged. The plagiarism only came to light during the manual search through all last accepted student answers.  This case underscored two things: (a) the critical role of TA diligence and thoroughness -- an automated flag on a few people was only the symptom of a bigger problem; solution analysis and subsequent manual checks were necessary -- and (b) the effectiveness of interviews in confirming suspicions.  In this instance, even without having seen the MOSS report, the interview process successfully determined which students had copied code and which had genuinely taught themselves in a ``whiteboard session'' and independently written the solution based on that.  It was, however, a sobering reminder that our workflow for reviewing MOSS findings is far from a complete picture of student plagiarism in a course\footnote{See \Cref{app:plag_ring} for a detailed account and ethical dilemma surrounding a \textit{large} cheating ring discovered in this process.}; clear plagiarism might have slipped by if not for the diligence of the TAs.

Another example of the failure of automation comes at the end of the S25 (and a handful of other) semesters.  At least two students were flagged for the last topic after all course activities had ended, including after the bonus contest -- the last opportunity for extra credit.  Due to simple instructor busyness in dealing with end-of-semester wrap-up, the cases were dropped.  There are other stories we could bring up (detecting code translated from one language to another is impossible with current automated tools, reviewing old student submissions after catching late plagiarism reveals massive plagiarism, students doing large numbers of new problems to pass the class after discovery of plagiarism, etc.), but we will end the story-based discussion of plagiarism cases here.

\subsubsection*{General Observations: LLMs}
The introduction of commercial genAI in 2022 added a twist to our observations. Starting Spring 2022, we began to suspect that some students were using GitHub Copilot (and later ChatGPT) to generate entire solutions. One clue was that a few submissions had a very distinct commenting style and verbosity that wasn't common among our students.  For example, elaborate function comments explaining trivial code, one comment for nearly every line, proper spelling, capitalization, punctuation, and spacing consistently in every comment\footnote{The easiest way to spot a solution exactly copied from ChatGPT is the ``capital with a space'' comment - a comment of the form \verb*|// Start comment| with a space and capital letter after the comment mark, a \textit{highly} unusual pattern to see in student code and an unseen pattern if replicated across the entire file.}
, which all felt off for student code.  We quickly learned (through our own active use of the tools for instructional purposes, see \cite{dickey2024gaide} for the framework we follow) that these are classic marks of ChatGPT and other LLM usage.

In one case, a C++ solution included library usage and language idioms far beyond what the student had shown they knew in other submissions - it used advanced CP-style code fragments that had never been discussed in class. When asked about it, the student could not adequately explain the code, eventually admitting they used ChatGPT ``to debug'' - an exceedingly common phrasing of students who blatantly copied from ChatGPT.  Interestingly, because this code was unique and not shared between students, neither Moss nor Kattis flagged it at all. The detection hinged entirely on the instructor's intuition and the subsequent interview.
This aligns with our earlier point: AI-assisted cheating often flies under Moss's radar, so catching it relies on anomalies in the work and verifying understanding. Since F22, we explicitly warn students that using AI to generate code is considered plagiarism in our course (unless specifically allowed for a task), and that they may be interviewed on any submitted code. This policy and warning may have deterred some misuse, but we remain vigilant. As AI tools become more integrated into IDEs and student workflows, distinguishing legitimate help from illicit outsourcing will be an evolving challenge in CP courses (and almost surely in CS education as a whole).  It has already become a burden in office hours, where students bring AI-generated code for TAs to debug.  Our policy for TAs is that if students don't understand what their code does or how they got there, start them from the beginning and rederive the solution theoretically without addressing their code at all.

One interesting observation happened when we performed the AI-Lab intervention, designed to teach educationally productive AI use without becoming overreliant \cite{dickey2025evaluating}, from S24 through S25.  In S24 and F24, students were allowed to use GenAI however they wished after topic 8 (the intervention topic), but in F24, GenAI had evolved enough since S24 that we could not assign problems hard enough that (a) it could not do them and (b) we could reasonably expect CP1 students to be able to do them.  In S25, we changed the allowance to only use it for one topic and made the problems much harder than average, such that it could not do most of them but could give interesting insights into the solution (and thus harder than an average CP1 student could do on their own). The AI-Lab topic for all 3 semesters had 5 problems scaled down to 4 problem's worth of points to acknowledge the benefits and helpfulness of GenAI in CP.  We found this intervention helped students greatly learn how to teach themselves things they didn't know using genAI, greatly reducing office hour and EdStem load.

\subsubsection*{Concluding Observations}\label{sec:experiences:conclusions}
Finally, through redesigning the CP sequence, the author implemented a mix of preventative and detective measures that yielded some positive results. Preventatively, we diversified the problem sets, and switched to Kattis from HackerRank (much better admin interface, plagiarism detection, test case quality, and overall problem quality), and sometimes personalized aspects of problems (for example, using different input data or variants in CP3) to reduce wholesale solution sharing.
We furthermore made it routine to run Moss (and check Kattis's similarity tool) after each week's assignment and promptly review the top matches. Those that looked suspicious were followed up with an interview focused on that assignment (in addition to regular interviews). In nearly 100\% of the cases, the suspects faltered in explanation of their code or directly confessed, confirming the plagiarism, indicating we were erring on the side of caution. Interestingly, suspect students rarely continued cheating after being caught once - the combination of detection and an academic integrity consequence (usually a zero and threat of report next time) had a strong deterrent effect on repeat offenses\footnote{Not counting the 2-3 extreme cases per 2 semesters of students caught at the end of the course and post-processed to find they had cheated the entire semester; these are always reported.  Without completing one make up problem for every plagiarized problem (usually in the last 3-4 weeks of the semester, an extremely high requirement), they failed the course. The underlying philosophy is ``if you're willing to prove to me you know the course material, I'm willing to give you a chance to do so.''  Around 30-40\% of these cases fail to complete and fail the course.}.

In summary, the on-the-ground experience from CP1 highlights that no single method catches all cheating. We have seen blatant copying that tools caught easily (provided we have time to look at the reports), subtle copying that only interviews exposed, and AI usage that neither tools nor peers would reveal directly. The experience reinforces findings from broader research: a high-precision approach (few false accusations) means many clever cheats will slip through \cite{chen2018cantstop}, whereas a high-recall approach (catch everything) tends to over-flag honest work \cite{hicks2024approach}. Balancing these is tricky. The stakes in a competitive programming course - where students build skills for competitions like ICPC - are such that we want them to truly learn, not just meet some arbitrary standard in a setting not applicable to the real world. This is why CP courses around the world rarely, if ever, have exams; the focus is on practice and practical skill growth. In the next section, we analyze in depth the detection methods used, evaluating how each contributed or failed to address the plagiarism cases observed.

\section{Analysis of Current Detection Methods and Their Limitations}\label{sec:currmethods}
Effective plagiarism detection in competitive programming requires a combination of tools and techniques. Here we dissect the main methods currently in use - automated code similarity tools (Moss and the Kattis system), visual/manual review of code, and interactive suspect interviews - drawing on both the literature and the CP1 course experience to evaluate their strengths and weaknesses. We also analyze how these methods hold up in the face of new threats like AI-generated code. \Cref{tab:summary_strengths_weaknesses} at the end of this section provides a comparative summary of the key points.

\subsection{Code Similarity Tools (Moss, Kattis, etc.)}
\subsubsection{How they work}
Code similarity tools like Moss and Kattis's plagiarism checker are built on algorithms that can identify segments of code that are identical or highly similar between two or more programs. Moss, for example, tokenizes the source code (converting it into a sequence of language tokens) and then applies a winnowing algorithm to compute a set of fingerprints for each file. It then compares fingerprint sets between all pairs of files to find matches above a certain length. The output is a list of pairs with a similarity score and excerpts of the matching code regions \cite{schleimer2003winnowing}, see practical examples in \Cref{app:MOSS}. Kattis's system uses hash comparisons and n-gram analysis \cite{Olsson_Norlin_2020}, which conceptually is quite similar (n-grams of tokens serve as fingerprints), to compare a submitted solution to their entire worldwide history of submissions to that problem (including GitHub solutions). These tools can ignore differences in whitespace, comments, and even sometimes variable names (since those changes don't affect the token sequence structure much; Moss is better at indicating this than Kattis, anecdotally). They are language-aware to an extent and usually support multiple programming languages, but do not by default compare solutions in different languages.

\subsubsection{Strengths}
Automated tools are fast and scalable. They can handle tens of thousands of submissions and compare them all against each other in minutes - a task that is impossible to do manually. They are also systematic and unbiased in their coverage: every pair is checked, not just those one might suspect. This thoroughness often uncovers plagiarism networks that instructors weren't even aware of (e.g., discovering that student A copied from student B who copied from student C, etc., by chaining similarities). Moss, having been used for decades, has a proven track record and is sensitive enough to catch cases with only light disguise. In many studies and anecdotal reports, Moss or similar detectors catch the vast majority of direct copy-paste or lightly modified cheating instances (levels 1-3 in the Faidhi-Robinson scale \cite{faidhi1987empirical}) with high reliability. They are excellent for an initial triage: Moss can \textit{``sav[e] teachers a lot of time by pointing out the parts of programs that are worth a more detailed examination''} \cite{schleimer2003winnowing}. In our CP1 use, viewing Kattis's automatic, worldwide, built-in checker after each assignment and running Moss on our class's submissions plus some ChatGPT and GitHub solutions consistently flagged the most egregious collusions (often confirming suspicions or revealing pairings we hadn't noticed). The tools integrate well into an auto-grading pipeline - for instance, Kattis automatically showing similarity results to TAs ensures we don't forget to perform a check.

\subsubsection{Weaknesses}
The primary weakness is over-reliance on superficial similarity. As Moss's own guidelines emphasize, it cannot determine plagiarism by itself - it only finds similarities, which might be legitimate or illegitimate. In competitive programming tasks, as discussed, legitimate similarities are common due to standard solutions, data structures, and algorithmic pieces (e.g. code pulled from course slides, the textbook GitHub, and FastReader). Thus, instructors must spend effort separating true vs false positives in the output. Setting a threshold is tricky: too low and you waste time on normal code resemblances; too high and you might miss more cleverly disguised cheating.  See step 1 of procedure in \Cref{list:plag_procedure} for details on how we worked through those issues. Another weakness is evasion by obfuscation. If a student is determined, they can apply transformations to the code that reduce the detectable similarity without too much effort - for example, changing the structure of loops, breaking one line into several, adding irrelevant computations, or reordering helper functions. These changes, especially in combination, can drop the similarity score significantly while the core logic is still stolen. Research shows that automated obfuscation scripts can defeat older detectors: e.g., a study demonstrated that inserting/reordering statements (using a tool) could fool Moss to the point that many plagiarism pairs fell below its detection threshold \cite{sauglam2024obfuscation}. Although newer tools like JPlag 5 have improved in this regard, no tool is fully obfuscation-proof. A savvy cheater might even test their modified code through a publicly available similarity checker to see if it would get caught, iterating until it passes (though Moss itself is not open to students, there are similar open-source checkers they could use).

Another limitation is that these tools usually only compare submissions within the given set (and possibly a provided base code set). They don't inherently have a database of all code on the internet. Kattis has the advantage here by comparing against every previous submission worldwide, which means if a student copies code from a solution someone had submitted to Kattis in a prior semester, at another university using Kattis, or to the public problem available on \href{open.kattis.com}{open Kattis} (like solutions posted to GitHub), it could be caught. But if the source is completely external (e.g., a solution generated fresh by AI), it won't be in the database. Instructors can manually add known external solutions to Moss as ``base files'' to check against, but this relies on the instructor guessing where students might copy from. In CP1, we did this: we would include around 5 GitHub solutions and 3-5 ChatGPT solutions in the Moss run. This helped catch quite a few cases of students copying from GitHub and ChatGPT. Unfortunately, however, we cannot preemptively include every possible source. This is why AI-generated code is such a blind spot for similarity tools - it's new code not in any repository (and if multiple students independently ask ChatGPT the same question, their outputs might still differ in implementation details, since responses vary by company, model, and even time of day).

Lastly, these tools have little understanding of semantics. They don't truly ``understand'' what the code does; they look for matching text/token patterns. Two codes that do the same thing with completely different implementations will not be flagged. For instance, one student might implement an algorithm iteratively, another recursively - if one copied from the other but painstakingly rewrote it in a different style, Moss will see very low textual similarity. More advanced semantic plagiarism detectors exist in research (using abstract syntax trees, greedy string tiling, or even program behavior analysis \cite{paiva2022assessmentsurvey,bejarano2015detection}), but they are not yet mainstream in classroom use. The typical Moss/Kattis approach focuses on surface similarity, which is both its strength (speed) and its Achilles' heel (vulnerable to deeper changes).

In summary, Moss and similar tools are indispensable for initial detection, catching the low-hanging fruit of plagiarism efficiently. However, they fail to detect cases where similarity is either obfuscated or inherently low (as with AI outputs). They also over-detect in scenarios of convergent solutions, requiring human interpretation. Their proper use thus demands a complementary method - usually a manual review or confirmation step - to make final determinations.

\subsection{Manual Visual Review}
\subsubsection{How it works}
Manual review involves instructors or TAs reading through code submissions (or diffs of two submissions) to look for signs of copying. This can be done proactively (reading each submission for style anomalies or known patterns) or reactively (examining specific pairs flagged by an automated tool or suspected by other means like plagiarism on later assignments). Reviewers look at aspects like structure, coding style, variable naming, comments, and even formatting. Often, when two pieces of code are side by side, plagiarized sections become obvious - e.g., a long sequence of operations that match in both, especially if they contain idiosyncratic elements (the same unusual constant, the same typo in a comment, etc.). A careful reviewer will also consider the context: do the two students share a lab or friend group? Have they submitted similar code in other assignments? Did one submission come shortly after the other? All these pieces can form a case for or against plagiarism. In our course, manual review typically came into play after getting a Moss report: for each pair above a similarity threshold, a TA would open both codes and compare them to verify if the matches were due to common boilerplate or truly suspicious copying.  Pairwise comparison isn't always necessary.  For the experienced instructor or TA, detecting plagiarism can come down to unusual submission patterns, code style, language switching, unusually good or unique solutions, and similarity to past submissions.  In extreme cases, all previous submissions of one student are checked, with each confirmation increasing analysis of other submissions\footnote{This can be come onerous with 44 total problems in the semester plus 7 bonus problems (A is 40, Pass is 26, 1 point per problem, lots of bonus available).  The worst case we had was in F23, one student plagiarized 16 problems out of the 29 they submitted for credit. They submitted 16 carefully checked make up problems to barely pass the class.}

\subsubsection{Strengths}
The greatest strength of human review is flexibility and insight. A human can notice when two pieces of code ``feel'' the same logically even if they aren't textually identical. For example, if two students implement the same novel algorithm with different variable names and order of substeps, an experienced reviewer might still recognize the algorithmic similarity beyond whiteboard similarity - something beyond current automated tools' reach. Humans can also pick up on semantic errors or uniqueness: if two submissions share the same subtle bug (say, off-by-one error in the same corner case) or the same creative approach to an optimization, that's a possible signal of illicit code collaboration or copying that a raw text match might miss. Moreover, a person can incorporate external information - knowing a student's coding style or capabilities. We've had cases where simply knowing that Student X usually writes in a very simple, novice style and suddenly a submission came in (on Kattis or in office hours) with highly sophisticated code prompted a manual check (which then revealed the code was likely copied).  An automated tool wouldn't flag that because it's looking at similarity across students, not complexity or authorship consistency.  The reverse is also true - a student who regularly submits at a high-quality, cleaned up level or with a advanced coding style can help clear them if flagged for an unusual or advanced technique or style.

Manual review is also how false positives are cleared. As noted earlier, automated systems may flag a lot of code pairs that aren't actual plagiarism. Only by reading them can one confirm. For instance, Moss might highlight that two solutions both have a loop and an if-statement in similar order - but on inspection, one sees that's just because both are implementing a common pattern, not that one copied the other. A person can quickly dismiss such a case as benign. The Moss creators explicitly state the need for human judgment: \textit{``it is still up to a human to look at the parts of code that Moss highlights and make a decision about plagiarism''} \cite{schleimer2003winnowing}. The process in our course typically involved marking each flagged pair as either ``Too Weak'' (similarity is coincidental/acceptable) or ``Suspect'' (requires further action), which only an instructor or experienced TA could do accurately.

\subsubsection{Weaknesses}\label{sec:currmethods:manual:weaknesses}
The main weaknesses are scalability and consistency. In a large competitive programming class, manually reviewing every submission or every pair is infeasible. Even beyond 20-30 students, scanning only last accepted solutions linearly (not $n \choose 2$) to notice unusual or obvious issues can consume up to 1-1.5 hours a week per problem.  Furthermore, even reviewing all flagged pairs can be daunting if many false positives occur. Humans also get fatigued and can miss things, especially when looking at a lot of code in a short time. There's a risk of bias - an instructor might scrutinize a student they suspect more harshly than another, or conversely give a favorite/honors student the benefit of the doubt consciously or unconsciously. To mitigate this, it helps to have multiple reviewers or some anonymization, but that's not always practical. Another weakness is that manual review is post-facto and reactive; it generally detects plagiarism after it has occurred (and after the assignment is submitted). It doesn't inherently prevent it. And unless interviews or further investigation follow, a student could still deny wrongdoing (``we just thought alike'') leaving the instructor to make a tough judgment call. In borderline cases without a ``smoking gun'' (e.g., not 100\% identical code or eerily similar enough structure), instructors might be hesitant to formally charge a student with academic dishonesty. This is why some departments err on only punishing the most blatant cases - to avoid wrongful accusations - which in turn might let some clever cheaters escape consequence. (It is worth mentioning here again that students working together without sharing code usually submit vastly different solutions, in our extensive experience pursuing plagiarism - see discussion in \Cref{sec:background:genai}.)

In CP1, we found manual inspection most valuable as a verification step and for catching a few tricky cases the tools missed. But doing it extensively required motivated TAs and time.
Manual review, while effective, can become a bottleneck or single point of failure if not consistently applied.

\subsection{Interviews and Suspect Interviews (Authorship Verification)}
\subsubsection{How it works}
When plagiarism is suspected or as a routine part of assessment, an instructor can hold an interview or oral exam with the student. This can range from a casual conversation about how they solved the problem, to asking them to explain parts of their code, to live-coding a small variant of the problem. Usually, we do not let students see their code for reference, only the problem statement (problems are short enough it shouldn't matter for concepts and solutions; clever cheaters can learn from their copied code quickly to answer questions and appear as if they know the solution well).  The goal is to confirm the student's understanding and authorship. In suspect cases, interviews often occur after something has flagged the student (like a Moss report or an anomalous submission). In our course, besides scheduled mastery interviews, we would call in specific students for an extra interview if their work was under question. The interview typically starts with general questions (``Explain how your program works'', ``Why did you choose this approach?'') and might escalate to pointed ones (``I notice your code uses an idiom we didn't cover - where did you learn that? Could you implement it differently?''). See \Cref{protocol:interviews} for our specific protocol.  Often, just requiring the student to reproduce a small part of the code or solve a simpler analog on the spot is very telling - an authentically prepared student can usually do this, whereas one who copied might panic or produce a very different-looking solution.

\subsubsection{Strengths}
Interviews provide a direct evidence of understanding (or lack thereof). If a student wrote the code, they \textbf{\textit{will}} be able to explain the reasoning behind it, even if they don't recall every line perfectly or the reasoning was minorly flawed. If they can't explain it at all, that strongly implies the work was not genuinely theirs. This method transcends the code itself - even if the code was completely original or heavily obfuscated (thus fooling tools), the student's inability to articulate it gives them away. Interviews also have a powerful deterrent effect. Knowing that an interview is possible (or guaranteed) makes students think twice about cheating.
Furthermore, interviews turn the plagiarism check into a learning opportunity: sometimes a student might have borderline collaborated or gotten more help than appropriate, but in preparing for the interview they end up studying the solution deeply, thus learning the material after the fact (two very powerful metacognitive strategies: reviewing one's own work and explaining it to others). While this doesn't excuse the initial dishonesty, it does ensure the student doesn't move forward entirely ignorant. In a roundabout, ironic way, the student is meeting some of the Learning Objectives this way.  In a few cases, we've had interviews where a student initially didn't understand their own submission (raising our suspicion), but as we guided them through questions, they improved their understanding - such cases might result in a grade penalty but still a better educational outcome than if we'd just failed them without discussion.  In this case, the interview is aligning very closely with how we run office hours, making the student come to the right conclusions with a guided framework without giving away any answers.

Interviews are also adaptive. An interviewer can probe specific unusual parts of the code (``What does this section do? Why is it needed?'') or ask the student to solve a related problem to see if they display similar coding skill. This flexibility makes it hard for a student to fool the process unless they truly have learned the material. It's often in interviews that confessions or revelations come out - faced with pointed questions, students who did cheat usually admit it (especially if they realize the alternative is to keep lying and potentially face a harsher penalty if found out). Thus, it's a great method to get to the truth of ambiguous cases.  In practice, our success rate approaches 100\% in detecting plagiarism this way.

We finally find that the title ``interview'' brings some ease to student perceptions.   Interview preparation is marketed as part of the point of CP, and practicing explaining their code and algorithmic ideas to an expert is highly beneficial in preparing for them.  In reality, these interactions are oral exams.

\subsubsection{Weaknesses}
The main downside is the resource demand. Conducting interviews for many students is time-consuming, as discussed. It also requires a certain interviewer expertise to be effective. A novice TA might not know how to corner a suspect who is evading answers, whereas an experienced professor might spot inconsistencies quickly. So, training and consistency in interviews are concerns. In our CP1, only the instructor and head TA are allowed to conduct them, and the head TA shadowed the instructor for several hours to learn what to look for. Furthermore, an undergrad should never be solely responsible for failing a student, so the head TA says something like, ``Thanks for coming in, we'll get back to you later this week.''

Additionally, interviews could introduce stress and subjective evaluation. A student might genuinely know the material but have an off day (nerves, etc.) and perform poorly in an oral assessment, potentially leading to a false suspicion. Clear protocols help (for example, what set of conditions must be met to fail a student interview). Another consideration is fairness: if only some students are pulled aside for suspect interviews (based on a tool's flag or instructor hunch), could there be an unconscious bias in whom we target? We try to mitigate this by having interviews with everyone - 10 students per week, everyone is guaranteed to be selected, otherwise random (so it's not only ``the suspected cheaters'' who face oral exam, which could be seen as accusatory) - and find this works well; the Moss- and Kattis-flagged cases that are suspicious but not solid are also called in (weak/unsure cases are fairly rare). That said, inevitably more time is spent on those suspected, which is appropriate but one must ensure those decisions are based on evidence, not profiling. 

In terms of failure modes for interviews: A particularly prepared cheater might anticipate questions and rehearse explanations (for instance, if they got the code from a friend, that friend might coach them on how it works). In a few rare situations, this could fool an interviewer initially. However, it's very difficult to maintain a lie under detailed questioning without cracks showing, unless the student has basically learned the code to the point it's indistinguishable from having written it. At that stage, ironically, the educational harm is partly mitigated (they learned something, even if by a convoluted route). Another potential issue is scalability when classes are huge - it simply may not be feasible to interview everyone in a 300-student class without a fleet of TAs.  On the other hand, we found that having 2 people available was good enough for 10 interviews a week * 5-10 minutes on average $\approx 75$ minutes a week for 12 weeks.  In total, 120 interviews over 12 weeks.  Adding 2 GTAs or qualified UTAs would approach 250 interviews; 4 dedicated office hours a week for around 250 students being interviewed total is in the realm of feasible solutions.

Despite these challenges, our experience and the general sentiment in recent educational discussions indicate that some form of interactive assessment is increasingly necessary. With AI able to do homework, only oral or supervised practical exams might reveal who actually has the skill. Thus, interviews or viva voce are regaining attention as a serious assessment tool, not just an ad-hoc plagiarism check \cite{dawson2024validity,dawson2025How}.

\subsection{The GenAI Factor: Why Current Methods Are Faltering}
Each of the above methods comes under additional strain due to genAI. Let's briefly articulate this for completeness:
\begin{itemize}
    \item \textit{Similarity tools vs AI:} As noted, Moss and Kattis largely cannot flag AI-generated code because it doesn't closely resemble other submissions or known sources \cite{taylor2023plagiarism}. Typically, AI answers are typically considered unique enough to not resemble each other, but we have found that generating 4-5 total solutions from 2-3 sources catches a decent percentage of people.  This could, in part, be due to most of the prompts students use being dominated by the problem statement, thereby making many student prompts appear similar and produce similar results. Unfortunately, if a student doesn't copy paste (or something similar), they can use AI and likely not appear in any similarity report - a big gap in the safety net.
    
    \item \textit{Manual review vs AI:} A human might notice oddities in AI-generated code, but as AI improves, its style might become more indistinguishable from a student's. In fact, if a student prompts the AI with ``write in a simple style'' or then refines the AI output, it could look quite ordinary. Manual graders might have a hunch (``this seems too advanced/perfect''), but they may not have concrete proof. Moreover, as of right now, the code genAI produces is written around the average quality of an undergraduate CS student; Without a known source to compare to, one can only quiz the student or look at their wider coursework for consistency.  On the other hand, in S25 we found that genAI created solutions that were higher quality, more creative, and better written for a few of the problems, allowing us to distinguish not by what was bad or unusual but by what looked ``too good.''
    
    \item \textit{Interviews vs AI:} Interviews remain the strongest defense. If a student used AI but doesn't understand the solution, the interview will expose that. If they do understand it (perhaps by studying the AI output), then at least they learned the content, though there's an argument about the ethics and lost practice of having not written it themselves. There is an edge scenario: what if a student uses AI to generate code and then thoroughly learns that code? In an interview they might pass as if they wrote it. In such a case, academic integrity was still breached (they didn't create the solution), but our detection might fail because the student has reverse-engineered their AI-given answer. This scenario could become more common as students realize they need to defend their work. It presents a philosophical and pedagogical question: if they truly understand the solution now, is the main harm just the unfair advantage on the assignment? (They still deprived themselves of the problem-solving practice, which might show up later though.) Regardless, interviews would not catch such a student because from an oral standpoint they are competent. This is akin to a student who copied but then studied so well that they can explain everything - rare, but possible.  It also brings into question the purpose of and reasoning behind the Learning Objectives of CP -- which argue that the actual problem solving process is what we teach, and therefore shortcutting this process deprives them of future problem solving ability.  Pedagogically, we know that students fail to generalize problem solving processes as well when using the \textit{understand} and \textit{apply} cognitive dimensions instead of the \textit{create} dimension from Bloom's Revised Taxonomy \cite{krathwohl2002revision}.

    \item \textit{Impact on incidence:} As the UIUC study showed, genAI's availability correlated with an increase in cheating frequency \cite{chen2024plagiarism}. Students who might have resisted cheating due to fear of getting caught or lack of an easy source might now be tempted since AI is easy and ``safe'' under current policing. This means instructors are facing not a static problem but a growing one, potentially overwhelming the current detection setup. \Cref{tab:num_plagiarism_incidents} shows part of this picture, to which we (anecdotally) add that the majority of the rise in cases are concentrated at the beginning of the course).  We also have seen in a high rise (e.g., 13 of the 15 discussed in \Cref{sec:experiences:backtogeneralobsv}: Major Example) in students denying plagiarism and being less apologetic after getting caught (historically as low as 1 in 20 denied).
\end{itemize}

Having analyzed each method in isolation, it's clear that none is a silver bullet. The failures of plagiarism detection in competitive programming often arise from the cracks between these methods: e.g., an automated tool fails to flag a case which also happens to avoid obvious signs a reviewer would catch, and if no interview occurs, it slips through entirely. Or the converse: a tool flags too much, the staff can't follow up on everything, and some true positives get dismissed with the noise. In the CP1 experience, whenever a cheating case went unnoticed until after the course, it was usually because it wasn't flagged by tools and we didn't interview that student on that particular code - a gap in coverage.  Similarly, if it was late enough in the semester, the single choke point of instructor emails, punishments, and interviews caused some cases to be dropped -- a gap in coverage due to manpower.

\Cref{tab:summary_strengths_weaknesses} summarizes the key strengths and weaknesses of the discussed detection approaches, given the current context (including AI). This comparative view underscores why a multi-pronged strategy is needed.

\begin{table*}[ht]
    \centering
    \begin{threeparttable}
        \begin{tabular}{l|l|l}%
            
            Method & Strengths & Weaknesses\\ \hline \vspace{1mm}
             
            \makecell[tl]{Automated\\Similarity Tools\\(e.g. Moss, Kattis)} &
                \makecell[tl]{
                    - Scalable; quickly compares all submissions\\
                    - Effective for copy-paste or slightly modified\\(levels 1-3 \cite{faidhi1987empirical})\\
                    - Objective similarity metrics and highlighted\\matches for review\\
                    - Integrates into auto-grading pipelines \\(Kattis flags suspects automatically)
                } &
                \makecell[tl]{
                    - False positives on common or short solutions\\
                    - Evaded by obfuscation or rewriting; not robust\\to high-level code changes\\
                    - Limited to known sources: won't catch novel\\external or AI-generated code with no matches\\
                    - Requires human interpretation; similarity score\\alone doesn't prove plagiarism
                }\\ \hline \vspace{1mm}
            \makecell[tl]{Manual Code Review\\(visual inspection)} &
                \makecell[tl]{
                    - Human insight can recognize plagiarism beyond\\exact similarity (e.g. same logic or mistakes)\\
                    - Can differentiate between coincidence and\\collusion by considering context and style\\
                    - Essential for confirming tool results: can validate\\true positives and discard false ones\\
                    - Can reveal qualitative anomalies (sudden jump\\in code quality, unusual patterns) that tools ignore
                } &
                \makecell[tl]{
                    - Not scalable without significant time; labor-\\intensive to examine many codes/pairs\\
                    - Subjective and prone to error or bias; consistency\\depends on reviewer skill and experience\\
                    - Mostly reactive (detects rather than deters)\\and may still require confrontation to resolve doubts\\
                    - Can be overwhelmed if automated tools output\\too many candidates (triage problem)
                } \\ \hline \vspace{1mm}
            \makecell[tl]{Suspect Interviews\\(oral exams on code)} &
                \makecell[tl]{
                    - Highly effective at verifying authorship; exposes\\cases where students lack understanding of\\their own code\\
                    - Strong deterrent when students know they must\\explain their solutions\\
                    - Flexible questioning can adapt to each student,\\probing any suspicious aspects in depth\\
                    - Serves as a learning opportunity and can\\confirm genuine mastery for honest students
                } &
                \makecell[tl]{
                    - Very time-consuming; doesn't scale easily without\\many staff or random sampling of subsets of\\students per week\\
                    - Can stress students; performance may be affected by\\anxiety (needs careful handling for fairness)\\
                    - Relies on interviewer expertise or standardized\\questions (see \Cref{protocol:interviews}) to ask the right questions\\and interpret responses\\
                    - Targeted interviews (for suspects) might overlook\\some cases if initial suspicion isn't raised; requires\\combining with other signals\\
                    - Single choke point for final verdict and punishment
                } \\ \hline \vspace{1mm}
            \makecell[tl]{GenAI Detection\\(emerging challenge)} &
                \makecell[tl]{
                    - \textit{(Currently limited in capabilities - no robust }\\\textit{standalone tool yet)}\\
                    - Potential in code stylometry or anomaly\\detection: could flag if code style suddenly\\changes (e.g. indicating AI use)\\
                    - AI can sometimes be recognized through telltale\\signs or incomplete code that an instructor\\might catch
                } &
                \makecell[tl]{
                    - No reliable automated way to detect AI-written code\\by similarity, since it's original by design\tnote{a}\\
                    - Stylometric methods are experimental and can be\\fooled by rewrites or AIs instructed to mimic style\\
                    - Over-reliance on AI detection tools could lead to\\false accusations (as seen with text-based AI detectors)\\
                    - Essentially forces return to oral exams or proctored\\assessments to ensure student authored the code
                } \\
        \end{tabular}
        \caption{A comparison of plagiarism detection approaches in programming, with typical strengths and weaknesses. GenAI is included as a new factor undermining the others.}
        \label{tab:summary_strengths_weaknesses}
        \begin{tablenotes}
            \item[a] As of May 2025, \textit{all} AI-detection tools have either unreasonably high false positive or false negative rates, or their dataset for testing wasn't diverse enough to expose false negatives that show up in real life at the collegiate level, especially in computer science.  The authors strongly warn against institutional recommendations for usage of these tools in a way that directly impacts grades.
        \end{tablenotes}
    \end{threeparttable}
\end{table*}
\section{Discussion: Opinions and Suggestions for Improvement}
Our analysis clearly indicates that traditional plagiarism detection is failing to fully address the challenges in competitive programming education, especially under the onslaught of new cheating avenues. What, then, can be done to improve the situation? In this final section, we offer a perspective - grounded in both the discussed research and the author's first-hand experience - on how instructors and institutions might bolster their academic integrity defenses. These suggestions range from technical enhancements to pedagogical and procedural changes. While some recommendations may be aspirational, the intent is to spark further discussion and development toward more robust solutions.

\subsection{Adopt a Multi-Faceted Detection Strategy}
No single tool or method is sufficient anymore. Thus, it is essential that we combine approaches. This might seem obvious, but it requires deliberate course planning, dedicated TAs, and time, especially for courses larger than 15 people. For instance, an instructor could use Moss/Kattis for initial screening and routinely integrate short oral check-ins for all students (not just suspects). Even a 5-minute conversation per student once or twice in a term can establish a baseline of who likely writes their own code. Automated tools catch the low-hanging fruit, manual reviews of flags spot nuanced similarities, and interviews confirm authorship - together covering much of the cheating spectrum. The key is not leaving any one of these out entirely (except manual review of non-flagged submissions, see \Cref{sec:currmethods:manual:weaknesses}). Our CP1 course's success in catching issues was highest in semesters where we actively employed all three: running Moss on every assignment, scanning the results manually, and conducting interviews (both planned and as-needed). This multi-layered approach significantly raised the bar for cheaters - to get away with it, a student would have to evade the tool detection or produce nothing obviously odd for a human to notice, and somehow pass an interview, which is extremely difficult. Of course, this is resource-intensive; thus, the combination should be balanced with class size and resources. Large courses might implement a rotation (e.g., interview a random subset - say 10\% - of students per assignment, ensuring everyone is interviewed at least once in the semester) to distribute effort.

\subsection{Improve Tool Efficacy and Use Next-Generation Detectors}
While classic tools have limitations, research and development on better plagiarism detection is ongoing. Instructors and institutions should stay on top of these advancements and consider adopting them. For example, the enhanced JPlag 5.0 mentioned earlier offers greater resilience to obfuscation \cite{sauglam2024obfuscation}. If available, using such a tool in addition to or instead of the older Moss could catch cases that currently slip through. Another promising direction is dynamic or semantic analysis - tools that compare programs based on behavior or abstract logic rather than surface text \cite{cosma2011approach}. Some prototypes measure program similarity by running the code with various inputs (program behavior profiles) \cite{jhi2015programcharacterization} or by comparing graph representations of code (e.g., program dependence graphs) \cite{holma2012programdependencegraph, cheers2021academic, liu2006gplag}. These could identify when two programs produce identical outputs for all test cases in a suspicious way (usually not a problem - CP mostly requires identical output), or when the structure of algorithms is isomorphic despite different syntax (also not the most applicable since a good CP problem could have only 1 or 2 correct solution structures). While not widely deployed yet, as competitive programming tasks have clear specifications, such semantic checks could augment traditional methods.  This becomes particularly relevant if the course uses different input data or problem statement variations, as briefly mentioned in \Cref{sec:experiences:conclusions}.

Furthermore, we encourage leveraging data from the coding process itself. Modern development environments or online judge logs can record timestamps of edits, compilations, and submissions. Analyzing this data can reveal anomalies: for example, a student who typically struggles and makes many attempts suddenly submits a perfect solution in one go (with no intermediate tries) - this is a red flag. Hicks et al. (2024) used metrics like number of attempts and time to solve to detect cheating anomalies in CodeWorkout logs \cite{hicks2024approach}. We should incorporate similar analytics: maybe the online judge can alert instructors, ``Student X solved a normally difficult problem in 1 try with zero wrong submissions,'' which might warrant a closer look. Essentially, catching process anomalies (extremely fast completions, minimal edit iterations, large code diff in a short time span, frequently changing languages within one problem -- especially more than once or more than 2 languages, frequent compilation errors right before all test cases passed, etc.) can complement catching output similarities. Geniuses aside, a pattern of unexpectedly perfect first submissions often correlates with unauthorized help.

\subsubsection{Cross-Language Detection}
One of the largest failures in plagiarism detection for CP, in our opinion, is the complete inability of any commercially-viable tool to detect plagiarism across languages. A frequent case we caught when reviewing historic submissions of caught students was when they had taken a GitHub solution and either carefully translated it to another language or simply asked an LLM to do it.  These can be very blatant, down to identical variable names, and no tool we have used is able to detect it.  This is where symbolic code analysis could help detect plagiarism specifically in CP, due to common CP policies stating that any language may be used, but they all get the same time limit (as in real life).\footnote{The authors are very interested in such a tool and are open to collaboration in testing one in a large CP course (100 students), if one becomes available.}

\subsection{Integrate Code Stylometry for Authorship Verification}
A more futuristic but intriguing solution is to use AI for the opposite purpose - not to generate code, but to detect who likely wrote a given code. Research in code stylometry and authorship attribution suggests that programmers have unique coding styles, much like writers have distinctive writing styles \cite{mirza2015style,mirza2018style}. Features such as how one names variables, how they structure loops and conditionals, preferred idioms, and even common mistakes can form a ``fingerprint'' of an individual coder. If enough training data is available (e.g., a student's coding assignments over a semester; four problems per week quickly adds up), machine learning models could potentially flag a submission that doesn't match a student's prior style. In an academic setting, this could be used to detect contract cheating (outsourcing assignments) or AI usage. For instance, if a student's first 5 homework submissions were sloppy, poorly commented, and used only basic constructs, but the 6th is elegantly structured and uses advanced language features, a stylometry model would find it anomalous. Some preliminary tools in this domain exist (such as approaches presented in cybersecurity conferences to de-anonymize programmers \cite{caliskan2015anonymizing}). While deploying this in a course might require expertise and isn't plug-and-play yet, it's a direction worth exploring. It could serve as an additional signal for instructors: an automatic warning like ``This code is likely not written by Student Y based on stylistic analysis'' could prompt an interview or further check. We must note concerns: coding style can evolve, especially in students, and false positives are possible if a student suddenly improves or deliberately changes style (especially in consideration of use of office hours). So this would augment, not replace, other methods. Privacy and ethics are also considerations - essentially fingerprinting students' code needs to be done transparently and only for academic integrity purposes.

\subsection{Rethink Assessment Design in Competitive Programming Courses}
A preventative approach is to design course assessments that inherently discourage cheating or render it fruitless. In CP courses, completely removing standard problem-solving assignments isn't feasible - solving problems is the whole point. But we can introduce elements that require personal input. For example, after each coding assignment, have students submit a short reflection or explanation of their solution approach, mistakes they made, and bugs they had. This can be a few written sentences or a brief video. It's harder to plagiarize both code and a coherent explanation of why you wrote it that way. The reflection might expose mismatches (a student who copied code might not be able to articulate the strategy clearly or might unknowingly contradict the code). Another idea is to occasionally give individualized variations of a problem. For instance, if the main task is a known problem, tweak some parameter or require an extension that's unique per student or per small group (different data ranges, an extra feature, different output format, etc.). This makes direct copying less applicable (two solutions won't be identical if the requirements differ slightly), and even using AI becomes trickier if each student's prompt is different. However, this approach can be limited by fairness (ensuring variations are of equal difficulty) and auto-grading complexity.

Introducing more closed-book, live assessments can also help. Perhaps include a few in-class programming quizzes where students have to solve a small problem on the spot without outside help (e.g., exercises from Johan Sannemo's Principles of Algorithmic Problem Solving\footnote{Exercises from Principles of Algorithmic Problem Solving are available on Kattis at \url{https://open.kattis.com/problem-sources/Principles\%20of\%20Algorithmic\%20Problem\%20Solving}} \cite{sannemo2018principles}). This can calibrate their actual skill level and discourage those who rely solely on unsupervised take-home tasks for grades. Competitive programming courses might incorporate contest-style timed tests (with supervision) in addition to homework problems. If a student consistently performs much better on take-home assignments than in-class tasks, that discrepancy can be investigated. We note that in CP1, there is no space in the Learning Objectives for performing in timed contest settings (a bonus contest is held at the end of the semester, but is not required for students), see details in \cite{dickey2025removing} and an alternate perspective for adding competition (usually in CP3) in \cite{luo2025curriculum}.

One idea for a required final that is much closer aligned with CP-style assessment and also with the general trend in education research away from timed written exams is to hold a final ``technical interview,'' lasting no more than 15 minutes per person (split across the Instructor and Head TA, along with any other person trusted to correctly and effectively grade).  For 100 students, this amounts to 8 hours of work with 3 interviewers, 6.5hr for 4, and 5hr for 5.  Still bordering on unfeasible, but courses with more instructional staff may be able to accommodate this.

\subsection{Address the GenAI Issue Head-On}
Educational institutions are currently grappling with how to handle AI assistance. One approach is the honor code route - explicitly forbid using AI to generate solutions (as plagiarism). Another approach is adaptation - accept that students will use AI and instead teach them how to do so ethically \cite{dickey2025evaluating} (e.g., citing AI assistance, or focusing grading on something AI can't do like explaining rationale or modifying the AI output in creative ways). In a CP class aimed at skill-building, we lean towards restricting AI use, because the practice of solving problems is one of the Enduring Outcomes. Our suggestion is to clearly communicate to students that using tools like ChatGPT to write code for them is considered cheating (unless explicitly allowed). This policy should be in syllabi and discussed openly, including why it's in place (to ensure skill development via scaffolding and problem solving framework development). Alongside policy, we need to adapt detection: perhaps use AI to fight AI, as alluded to above. One could, for instance, run suspicious code through ChatGPT itself, asking ``How likely is it that this code was produced by an AI like you?'' - this is experimental and not reliable for formal evidence, but it might provide hints. OpenAI had an AI text classifier (now discontinued for low accuracy), and others are developing detectors; if any prove reasonably accurate for code, they could be used as one input in a holistic evaluation (with caution to avoid false accusations).  The author doubts this, however, based on the simple fact that when it comes to code, LLMs produce code that is the same quality or slightly better than undergrad CS majors (by definition, their coding style isn't good nor consistent yet, that is why they are in college).  As LLMs become better and produce better code, it is easy for a student to take a high-quality piece of code and make it worse by attempting to implement it themselves.

We also recommend faculty collaboration and data sharing regarding AI. If an instructor finds that certain AI prompts consistently yield a solution to a course assignment, that knowledge can be shared so others can check for telltale signs of that output. Perhaps in the future, plagiarism services might include cross-checking against a database\footnote{The authors do this manually by generating a set of genAI solutions and feeding them into Moss runs alongside student code, with decent success.} of known AI-generated solution patterns (if not exact matches, maybe certain distinct structures AI tends to produce). Until then, vigilance during manual review and interviews is the way to catch AI use.

\subsection{Encourage a Culture of Integrity and Mastery}
Beyond detection and punishment, it's important to address why students plagiarize and mitigate those motivations. In a mastery-based course, ideally students shouldn't feel the need to cheat - they can get help, take more time, or accept a small penalty and learn properly. We found that emphasizing the learning aspect and having regular check-ins reduced the temptation for many. Students knew they couldn't just fake it because eventually they'd have to prove themselves in an interview or the next harder problem. Additionally, openly discussing examples of plagiarism (anonymized) and their consequences in class can deter students. When they see that instructors take it seriously and can detect it, the risk of cheating may seem less worthwhile. Another angle is to make students aware of the personal consequences: not just disciplinary action, but the fact that if they cheat, they won't build the skill needed for competitions or advanced courses that this class prepares them for. In competitive programming, peers often pride themselves on skill - framing cheating as antithetical to that culture (you're only hurting your own ability to compete) can resonate.  Finally, a strong selling point of CP courses in general is the lack of tests and quizzes in favor of more practice.  Having to take a final can be a powerful deterrent.  Anecdotally, many students who were caught in the first two weeks, after finding out they were required to take a final, dropped the class (many more than usual; the rule was introduced in S25).

\subsection{Further Research and Continuous Improvement}
The opinions given here highlight areas for further inquiry. For instance, more research is needed on how to effectively integrate oral assessments without overwhelming instructors - perhaps intelligent tutoring systems could conduct basic code interviews with students using AI (with the AI flagging uncertain cases to a human instructor). Research can also explore how moderated and educationally-sound AI usage impacts learning longitudinally (presently, interventions like the AI-Lab \cite{dickey2025evaluating} have only been measured as cross-sectional studies).
Research can also explore how to formally detect AI-generated code with stylometry - maybe through analyzing randomness in variable names or patterns that statistically differ from human code. As this is a preliminary examination, we strongly advocate for the computing education research community to share findings on what works in this new landscape. Publications and case studies on plagiarism in the AI era (like the ones we cited in \Cref{sec:intro,sec:backgroundandlit}) are extremely valuable.

\section{Conclusion}
While the failure of plagiarism detection in competitive programming is a real and present concern, it is not a helpless situation. By doubling down on a combination of enhanced tools, smarter assessment design, and human-driven verification, educators can significantly reduce undetected cheating. The author's experience shows that even though a handful of cases might slip through in a given semester, the vast majority can be caught or deterred with a robust system in place - and importantly, students can be guided towards genuine learning instead of shortcuts. The landscape is evolving: genAI is the latest adversary to academic integrity, but it also challenges us to evolve our teaching and evaluation methods for the better. Going forward, competitive programming courses (and programming courses in general) will need to be proactive and innovative in preserving integrity. This might mean holding oral exams, more personalized tasks, adoption of new detection software, and an ongoing dialogue about the role of AI in learning. By implementing the suggestions above and continuing to share effective practices, the community can improve plagiarism detection and perhaps transform what was once an ``adversarial'' process into an integrated part of learning (for example, using interviews and reflections to deepen understanding, not just catch cheaters).

Ultimately, ensuring that students truly write and understand their code is paramount - not just to prevent cheating, but to fulfill the educational mission of computing programs. Plagiarism detection will likely never be perfect, but with rigor and creativity, we can shrink the gap between what students submit and what they genuinely know, even in the face of ever-changing technology and tactics.

\appendices %
\crefalias{section}{appendix}

\section{List of Tracked Kattis Solution Repositories}\label{app:trackedRepos}
We list the current set of Kattis solution GitHub repositories currently tracked and included in automatic plagiarism detection in CP1 and CP2 at Purdue University.  They are also the first cross-referenced when plagiarism is suspected but no matches were found.  This is by no means a comprehensive list of solution repositories, but it does cover many of the major offenders.
\begin{itemize}
    \item By far, the most used solution repo by students: \href{https://github.com/mpfeifer1/Kattis}{mpfeifer1}, and similarly: \href{https://gitlab.com/Syuq/Kattis}{Syuq}
    \item The second-most plagiarized source by CP students: \href{https://github.com/JonSteinn/Kattis-Solutions}{JonSteinn}
    \item Solutions for all of the online judges: \href{https://github.com/abeaumont/competitive-programming}{abeaumont}
    \item Kattis, HackerRank, GCJ, Codility: \href{https://github.com/mbobesic/algorithms-playground/tree/master/kattis}{mbobesic}
    \item Sorted by Competitive Programming 4 chapter: \href{https://github.com/BrandonTang89/Competitive_Programming_4_Solutions}{BrandonTang89}
    \item \href{https://github.com/RussellDash332/kattis}{RussellDash332}
    \item \href{https://github.com/dakoval/Kattis-Solutions}{dakoval}
    \item \href{https://github.com/Solopie/kattis-solutions/tree/master/java}{Solopie}
    \item \href{https://github.com/prokarius/hello-world}{prokarius}
    \item \href{https://github.com/minidomo/Kattis}{minidomo}
    \item Minor: \href{https://github.com/AnimalMother83/Kattis-solutions}{AnimalMother83}
    \item \textbf{\textit{Mega List}} for when we can't find others: \href{https://github.com/topics/kattis-solutions}{GitHub topic: kattis-solutions}
\end{itemize}
If those don't work and we suspect plagiarism, we also occasionally just google the solution.  One-off solution repos aren't uncommon.

\section{Plagiarism Ring: The Hustle}\label{app:plag_ring}
In the S25 mass plagiarism case (\Cref{sec:experiences:backtogeneralobsv}, \textit{Back to General Observations}), one student, codename \textit{Eve}, after being the most resistive to confessing plagiarism than any other student in the history of this course, was finally caught when they opened ChatGPT on their laptop. Not only did we find multiple solutions to assigned homework problems, we found solutions to a plethora of courses like CS251, CS253, CS240, CS242, sociology, polysci, history, english, psychology, and more.  In that ChatGPT history, there were \textit{hundreds} of chats within the last \textit{week} covering the breadth of nearly all approximately sophomore-level CS courses and Gen-Ed options.  It took 20 minutes just to scroll (at a decent pace) through one week of chat history.  The student then proceeded to \textit{brag} about the premium account (explicitly named something like \textit{Plagiarism Account}) and how they made their 12+ friends pay 2\$/month to use it.  The account costs 20\$/month: the student was bragging about hustling a cheating ring of 12 other CS students.  This student warranted special attention and was pointed out explicitly to the university in the mass report.

\subsection{Ethical Dilemma and Instructional Response}
During the interview with \textit{Eve} (which happened late at night after class, around 11pm in the sequence of interviews), the instructor obtained a list of collaborators only after giving three explicit assurances:

\begin{description}
    \item[\textbf{(A) Non‑investigation assurance.}]  
          The instructor would not report collaborators who were not enrolled in the instructor's own course.
    \item[\textbf{(B) Limited‑forensics assurance.}]  
          The instructor would stop scrolling through Eve's ChatGPT history after the single week already inspected and would refrain from collecting further screenshots or logs.
    \item[\textbf{(C) Ephemeral‑data assurance.}]  
          Once each name was verified as external to the course, the collaborator list would be deleted and no local copy retained.
\end{description}

\noindent By the next morning, the instructor realized these assurances conflicted with a broader professional duty:

\begin{description}
    \item[\textbf{Duty of care to the department.}]  Although Purdue's Office of Student Rights and Responsibilities does not mandate faculty reporting in every case, the University ``asks that you do so because it helps the university keep track of a student's disciplinary history and behavioral concerns accurately''\footnote{\url{https://www.purdue.edu/odos/osrr/academic-integrity/faculty-staff.html}}.  Withholding the list could therefore leave other instructors unaware of credible risks to the integrity of their assessments and prevent the university from responding to a pattern of incidents.

    Furthermore, reporting credible plagiarism risks to the department preserves fair grading across courses, equips colleagues to protect their assessments, and safeguards the program’s overall academic integrity.
\end{description}

\subsubsection{Moral Tension.}
Honoring the assurances risked enabling undetected plagiarism across the curriculum, undermining departmental grading fairness and ultimately eroding the quality of the department's degrees.
Conversely, an immediate report would break the instructor's word, something they hold in high regard, potentially discouraging future cooperation and eroding trust in restorative practices.  The decision space thus resembled the \textit{dual‑loyalty} dilemma familiar in medical ethics: personal commitments to an individual conflict with a fiduciary duty to the greater community.

\subsubsection{Resolution and Reporting Strategy.}
Guided by a consultation with one of the Assistant Department Heads for Computer Science, the instructor adopted the following approach that balanced policy compliance with the practical need to protect ongoing investigations:

\begin{enumerate}
    \item \textbf{University–level report (Student Rights and Responsibilities, OSRR).}  Only the students for whom direct, verifiable evidence existed -- \textit{Eve} and the one additional submitter in this class whose code matched the ``Plagiarism‑Account'' transcripts -- were included in the formal academic‑misconduct filing.  This satisfied the professional responsibility requirement while restricting disclosure to cases already supported by primary evidence.
    \item \textbf{Department–level alert (Instructor notification).}  A separate confidential memo, devoid of the reporting instructor's identity, was delivered by the CS department to instructors of the twelve students that the individual had shared access to a premium plagiarism‑enabling ChatGPT account with.  The memo recommended a submission review of previous assignments and caution moving forward.  Crucially, these names were not appended to the university report, preserving the distinction between confirmed misconduct and credible risk.
\end{enumerate}

\subsubsection{Confidentiality Constraints.}
A follow‑up meeting with Eve to explain this bifurcated approach proved impossible: the students were under active investigation in other courses, and any warning could have jeopardized evidence in those cases.

\subsubsection{In the End.}
In all, the duty to the department to preserve the integrity of the degrees and the department as a whole outweighed the individual instructor's responsibility to keep their word to the student.  We don't necessarily claim this was the fundamentally correct decision, but it was the decision that was made.

\section{MOSS Examples}\label{app:MOSS}
MOSS can give varied results for different types of problems.  For some, common algorithms are common sources of high similarity scores (e.g. Tarjan's for Strongly Connected Components), as students are allowed to retrieve such common algorithms from online sources (we strongly suggest the Competitive Programming 4 textbook GitHub\footnote{\url{https://github.com/stevenhalim/cpbook-code}}, and discourage GeeksForGeeks -- as we have found, over 4 years of crowded office hours, that $\ge 50\%$ of their code has subtle bugs that come out in the rigorous Kattis test cases -- but students may use whatever they wish as long as there is a link in the code).

\begin{figure*}
    \centering
    \includegraphics[width=\linewidth]{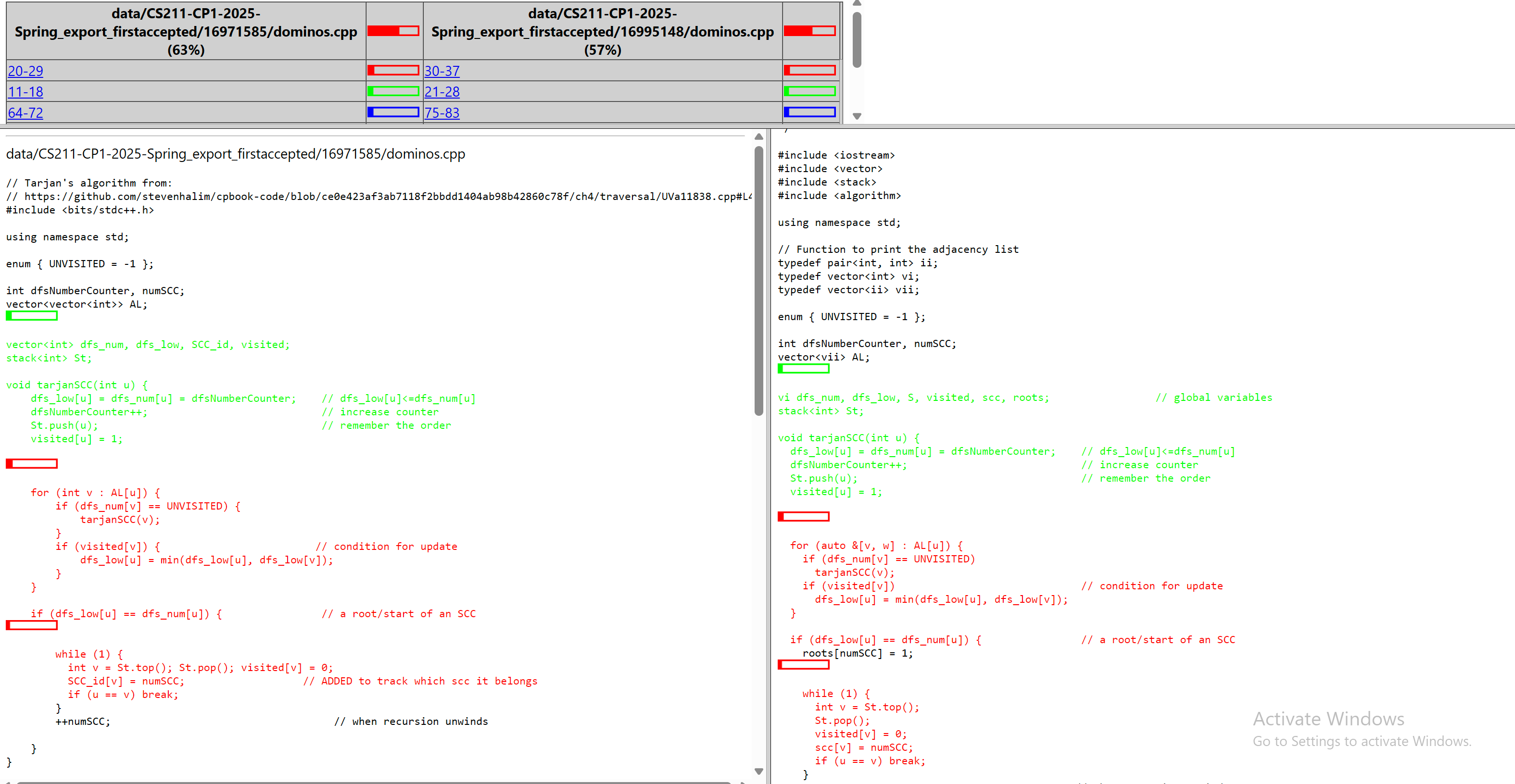}
    \includegraphics[width=\linewidth]{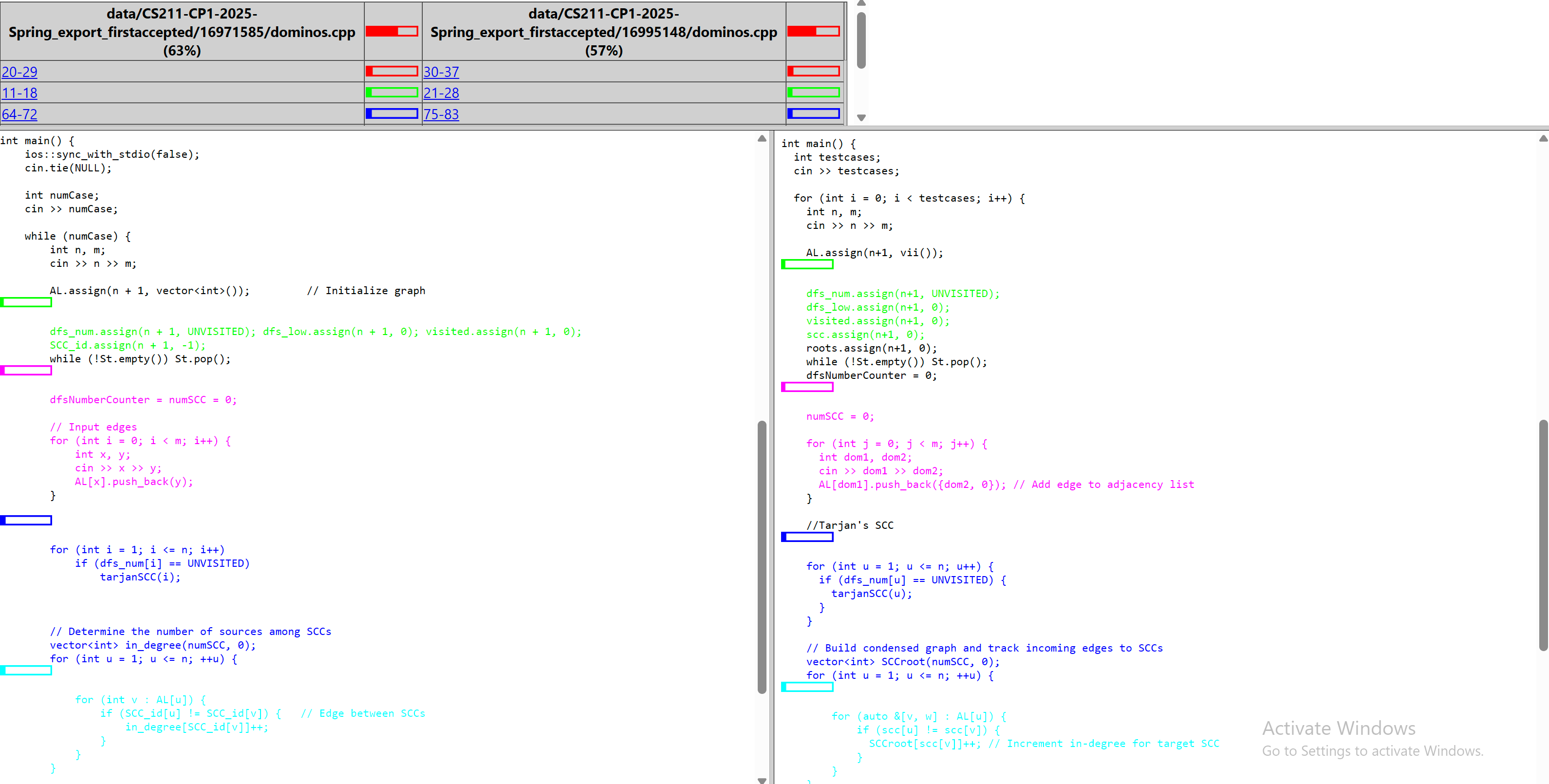}
    \caption{Example MOSS similarity score and output for Dominos, parts 1 and 2.  High similarity score due to common Tarjan's algorithm source.\\\url{https://open.kattis.com/problems/dominos}}
    \label{fig:dominos_p1}
\end{figure*}

\begin{figure*}
    \centering
    \includegraphics[width=\linewidth]{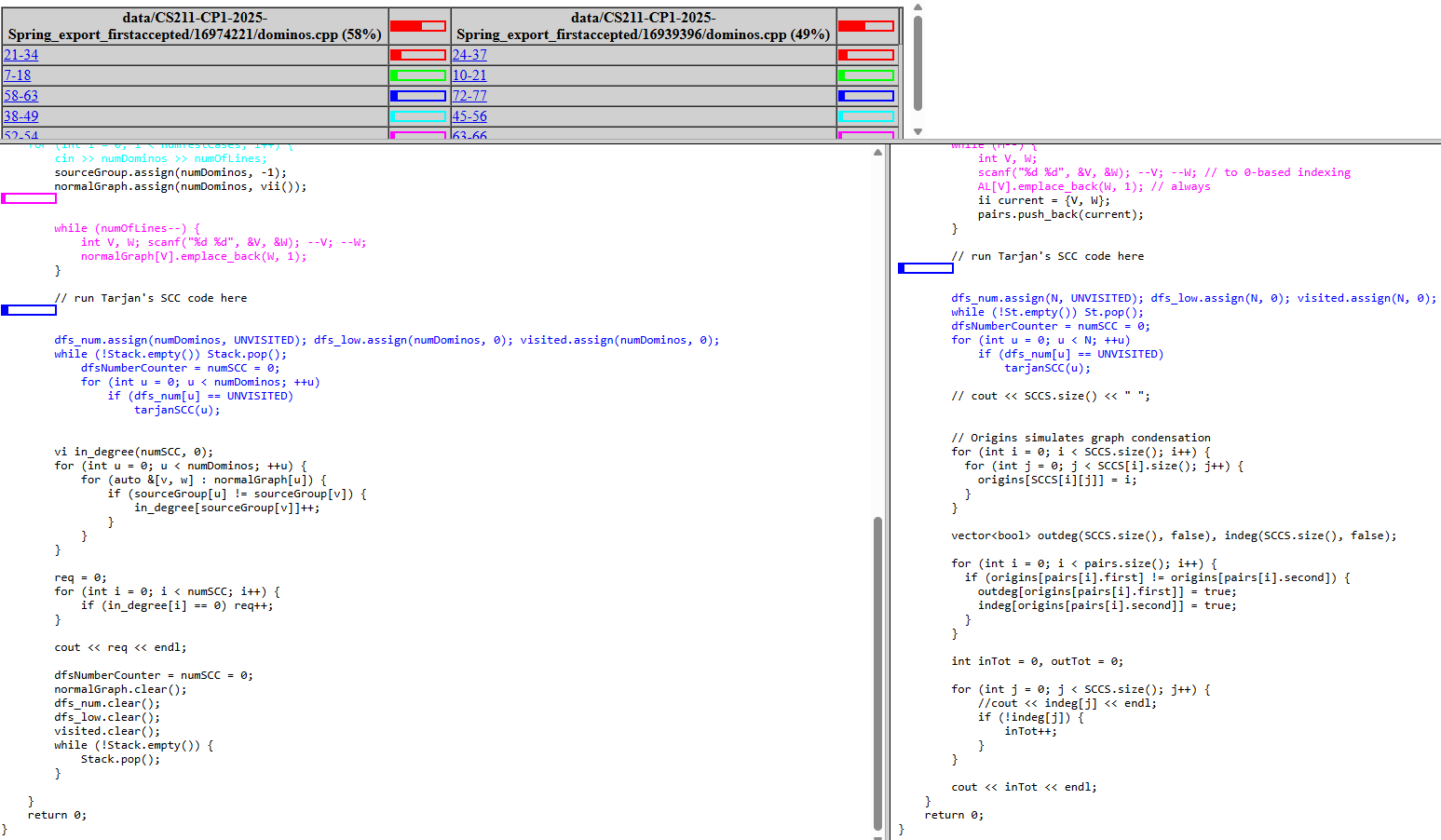}
    \caption{Example MOSS similarity score and output for Dominos, parts 1 and 2.  Great example of a high similarity score due to common Tarjan's algorithm source but drastically different solutions with clear student fingerprints (idiosyncrasies).}
    \label{fig:dominos_p2}
\end{figure*}

\begin{figure*}
    \centering
    \includegraphics[width=\linewidth]{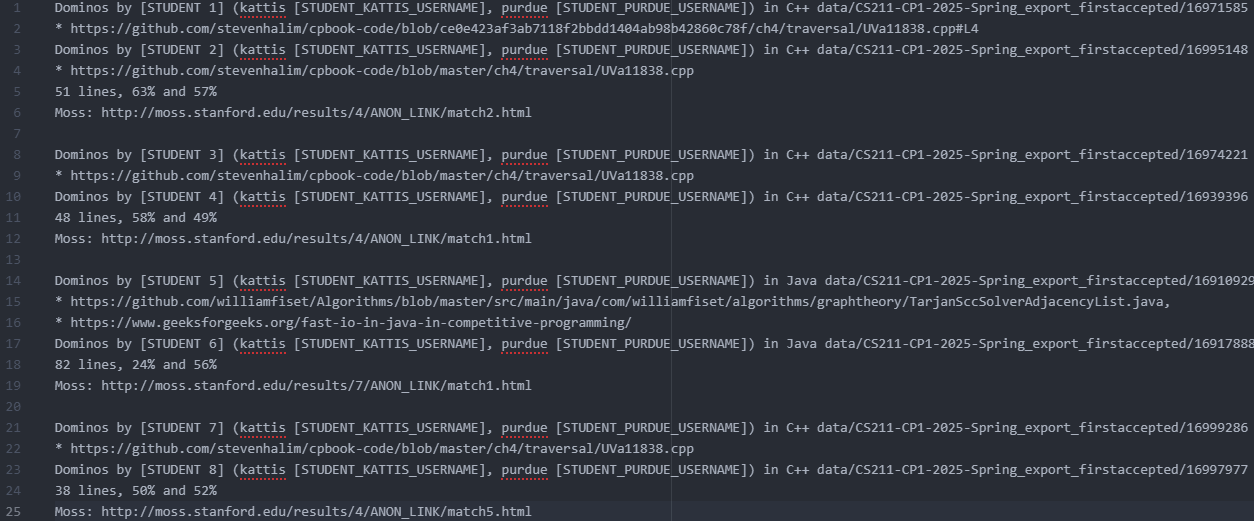}
    \caption{Anonymized MOSS Summary report, truncated.}
    \label{fig:moss_report}
\end{figure*}

\Cref{fig:dominos_p1} shows an example of a high-similarity score due to Tarjan's algorithm.  The fundamental code (how the students used Tarjan's) wasn't the same, and unique student fingerprints were easily detectable, but the similarity score was around 60\%.  A great example of two dissimilar solutions is showin in \Cref{fig:dominos_p2}, where the first student counts in-degrees of groups and the second uses graph contraction (``condensation'').  Both have clear ``student fingerprints'' -- the first in \verb|while !stack.empty(): stack.pop()| (instead of \verb|Stack = stack<>()|) and the second in (a) ``\verb|condensation|'' (very uncommon language, likely a misunderstanding by the student), (b) \verb|outdeg[origins[pairs[i].first]]| (a very convoluted way to go about what they wanted to do) and (c) commented-out debugging code. In each of those cases, an LLM would have produced cleaner code and a GitHub solution would have produced more unreadible and messy ``competition code.'' A truncated example of the MOSS summary report is provided in \Cref{fig:moss_report}.  The plagiarism TA for CP1 in S25 described cutoffs for manual review like this (at the end of the semester):
\begin{quote}
    It really depends on the difficulty of the problem... [A]ssuming a difficulty of Topic 6 and onwards [average open Kattis difficulty of 4.07 compared to the first half average of 3.51], I would look into all submissions with $>30\%$ similarities if it doesn't have like a shared algorithm. For the ones with a shared algorithm, I would look at the general trend and often start with 40 or 50. \\\emph{Segyul Park, May 2025, private message.}
\end{quote}
Topic 8 onwards were graph algorithms, a common source of high similarity scores.  This frequently acts as noise to hide plagiarism cases -- there was a $\ge 80\%$ reduction in plagiarism cases in Topic 8 and after.  This may also be due to the easier nature of the problems or the overall decreasing trend of plagiarism in the course as students either get more capable or more clever.


\section*{Acknowledgments}
A special acknowledgement goes to Pedro Sugiyama, the CP1 head TA, who routinely confirms and helps deal with plagiarism cases, amongst \textit{many} other miscellaneous activities.  We thank Segyul Park, the current TA assigned to detect and record plagiarism through MOSS and Kattis, among other things.  We also thank historical CP Head Ta Otavio Sartorelli de Toledo Piza and historical plagiarism TAs Joshua Yang (initiator of MOSS checking in CP1) and Thomas Marlowe.  Finally, we thank the rest of the past and current TAs for the CP1/2/3 sequence, as well as fellow instructors Arvind Ramaswami and Zhongtang Luo (CP2 and CP3).

The authors acknowledge the assistance of ChatGPT in the transcription and summary of initial notes on the paper and in final polishing/wording/clarity checks.

\bibliographystyle{IEEEtran}%
\bibliography{refs}%

\end{document}